\RequirePackage{fix-cm}
\documentclass[smallextended]{svjour3}       % onecolumn (second format)
\smartqed  % flush right qed marks, e.g. at end of proof
\usepackage{graphicx}
\journalname{Experimental Astronomy}
\begin{document}

\title{SARAS: a precision system for measurement of the Cosmic Radio Background and signatures from the Epoch of Reionization
}
%%\subtitle{Do you have a subtitle?\\ If so, write it here}

\titlerunning{SARAS: correlation radiometer for the Cosmic Radio Background}        

\author{Nipanjana Patra \and Ravi Subrahmanyan \and A. Raghunathan \and N. Udaya Shankar}

\authorrunning{Patra et al.} 

\institute{
	Nipanjana Patra \and Ravi Subrahmanyan \and A. Raghunathan \and N. Udaya Shankar
	\at Raman Research Institute, C V Raman Avenue, Sadashivanagar, Bangalore 560080, India \\
	\email{nipanjana@rri.res.in, rsubrahm@rri.res.in, raghu@rri.res.in,  uday@rri.res.in}
	\and
	Nipanjana Patra
	\at Joint Astronomy Programme, Indian Institute of Science, Bangalore 560012, India \\
	\email{pnipanjana@physics.iisc.ernet.in}
	\and
	Ravi Subrahmanyan
	\at
	National Radio Astronomy Observatory, PO Box 0, Socorro, NM 87801, USA \\
          Tel.: +1-575-835-7000\\
          Fax: +1-575-835-7027\\
	\email{rsubrahm@nrao.edu}
	}
\date{Received:  Nov 12, 2012 / Accepted: date}

\maketitle

\begin{abstract}

SARAS is a correlation spectrometer purpose designed for precision measurements of the cosmic radio background and faint features in the sky spectrum at long wavelengths that arise from redshifted 21-cm from gas in the reionization epoch.  SARAS operates in the octave band 87.5--175~MHz.  We present herein the system design arguing for a complex correlation spectrometer concept.  The SARAS design concept provides  a differential measurement between the antenna temperature and that of an internal reference termination, with measurements in switched system states allowing for cancellation of  additive contaminants from a large part of the signal flow path including the digital spectrometer.  A switched noise injection scheme provides absolute spectral calibration.  Additionally, we argue for an electrically small frequency-independent antenna over an absorber ground.  Various critical design features that aid in avoidance of systematics and in providing calibration products for the parametrization of other unavoidable systematics are described and the rationale discussed.    The signal flow and processing is analyzed and the response to noise temperatures of the antenna, reference termination and amplifiers is computed.  Multi-path propagation arising from internal reflections are considered in the analysis, which includes a harmonic series of internal reflections. We opine that the SARAS design concept is advantageous for precision measurement of the absolute cosmic radio background spectrum; therefore, the design features and analysis methods presented here are expected to serve as a basis for implementations tailored to measurements of a multiplicity of features in the background sky at long wavelengths, which may arise from events in the dark ages and subsequent reionization era.

\keywords{Astronomical instrumentation, methods and techniques \and Methods: observational \and Cosmic background radiation \and Cosmology: observations  \and Dark ages, reionization, first stars \and Radio continuum: ISM }
%\PACS{95.55.Jz \and 95.85.Bh \and 98.70.Vc \and 98.80.Es}

\end{abstract}

\section{Introduction}
\label{intro}

Following recombination at redshift $z \approx 1100$, the propagation of the cosmic microwave radiation (CMB; Dicke et al. 1965; Penzias and Wilson 1965) through the `dark ages' and `cosmic dawn' results in deviations of its spectral form from a Planck blackbody spectrum.   The early signatures arise owing to radiative transfer interactions of the CMB with the spin-flip 21-cm transition in the neutral hydrogen content of the universe, which depends on the spin temperature evolution of the gas.  Subsequently, in the post-reionization era, free-free emission from the ionized intergalactic medium as well as Compton scattering of CMB photons by the hot electrons in this gas imprint further deviations from the Planck form.  All of the above traces, however, constitute a small fraction of the total cosmic radio background (Jansky 1933; Clark et al. 1970) at long wavelengths: the relatively more recent history of star formation and AGN activity in galaxies cumulatively manifest as the radio source counts and constitute the bulk of the extragalactic radio background.  

Cosmological expansion and the consequent redshift results in a movement of the cosmological radiation signatures to longer wavelengths; therefore, precision measurements of the spectrum of the cosmic radio background (CRB) at long wavelengths is an important probe in modern cosmology.

Recent analysis of ARCADE 2 measurements (Fixsen et al. 2009 and references therein) of the cosmic radio background suggests that the extragalactic radio background is the CMB plus a power-law spectrum that has a brightness temperature of $(1.2\pm0.09)\times (\nu/{\rm 1~GHz})^{-2.60\pm0.04}$~K.   Surprisingly, the sky brightness corresponding to discrete radio sources detected in the deepest surveys to date account for only a fraction of the extragalactic radio background, even after excluding the CMB (Seiffert et al. 2011; Singal et al. 2010).  The identification of the extragalactic component in the ARCADE 2 measurements and its origin continue to be debated (Fornengo et al. 2011), with recent analyses and deep measurements of radio source counts ruling out most plausible populations of sources that might constitute the unexplained fraction (Vernstrom et al. 2011; Condon et al. 2012). Improved measurements of the radio background and, in particular, the spectrum at long wavelengths where errors are relatively larger, are important in estimating the spectrum of the unexplained part and thereby constraining the sources of this cosmic radiation. A recent improved measurement of the background spectrum at long wavelengths is in Rogers and Bowman (2008) and a new calibration method is discussed in Rogers and Bowman (2012).

Free-free emission from ionized gas in the post-reionization era adds to the radiation background, and is usually characterized by the optical depth to free-free emission $Y_{\rm ff}$ (Bartlett and Stebbins 1991).  Energy release at early times---for example, as a consequence of radiative decay or annihilation of relic particles---would be expected to manifest as distortions to a Planckian spectrum of the $\mu$ form (Burigana et al. 1991).  At later times, inverse Compton scattering of CMB by hot electrons of the intergalactic medium following reionization of the gas is expected to manifest as a `$y$'-type distortion of the CMB spectrum (Sunyaev and Zeldovich 1970).  Constraints to $Y_{\rm ff}$ are derived by fits of models for the evolution in the gas to measurements of the radio background; constraints on $\mu$ are derived from fits of spectra with a Bose-Einstein photon occupation number where the deviation from Planckian form is parametrized by $\mu$.  Deviations in the background brightness temperature at frequencies below about 1~GHz best constrain these terms and, therefore, provide a motivation for improved measurements of the long wavelength spectra of the extragalactic radio background.

Perhaps the key scientific motivation today for any measurement of the spectrum of the radio background at long wavelengths is the cosmological epoch of reionization (Shaver et al. 1999).  The evolution of the spin temperature of the gas, as the coupling between hyperfine level populations with gas kinetic temperature and radiation background changes with cosmic epoch, is expected to manifest as a distinctive signature in the radiation spectrum (Pritchard and Loeb 2008).  Significant deviations in the spin temperature of the gas from the temperature of the ambient radiation background commence early, at redshift beyond 100, as the  spin temperature is driven below the radiation temperature by the gas cooled by cosmological expansion.  However, this coupling effectively disappears by about redshift 30.  Spectral features associated with the formation of the first condensed objects are expected to appear at redshift about 20, when the coupling of spin to the gas kinetic temperature is renewed by the Wouthuysen-Field coupling (Wouthuysen 1952; Field 1958) induced by the Lyman-$\alpha$ from the first light.  Subsequent heating of the gas and reionization result in a transformation of the gas visibility from absorption to emission, and a diminishing of the gas emissivity in redshifted 21-cm as the neutral fraction vanishes; this final step in the reionization epoch is expected to occur progressively between redshifts of about 13 and 6 (Zahn et al. 2012).  

Reionization is expected to be effectively complete by redshift of 6 (Fan et al. 2006; Hinshaw et al. 2013), corresponding to an observing frequency of about 200~MHz, and hence all these features associated with the cosmological evolution of the gas is at lower frequencies.  The magnitude of the signal corresponding to the epoch of reionization is not expected to exceed 0.1~K and is at least three orders of magnitude smaller than the foreground.  As a consequence of the relative dominance of the foregrounds, any small instrument systematics would interact with the bright foreground and result in responses that may confuse the reionization signals.  It is therefore important to minimize systematics by paying careful attention to the instrumental design and characterization. 

The astrophysical evolution including the formation of the first compact objects, stars, galaxies, and the ionizing radiation from these sources are poorly constrained: therefore, precision measurements of deviations in the background spectrum from Planckian form are means to improved modeling of this era.  There are several ongoing and proposed experiments that aim to make precise measurements of the reionization signatures that are expected in the spectrum of the cosmic radio background: CoRE (Chippendale 2009), EDGES (Bowman and Rogers 2012), LEDA (Lincoln et al. 2012) and DARE (Burns et al. 2012); the instrument described here differs in many ways from the systems described previously.

\begin{figure}
\centering
\includegraphics[width=10cm]{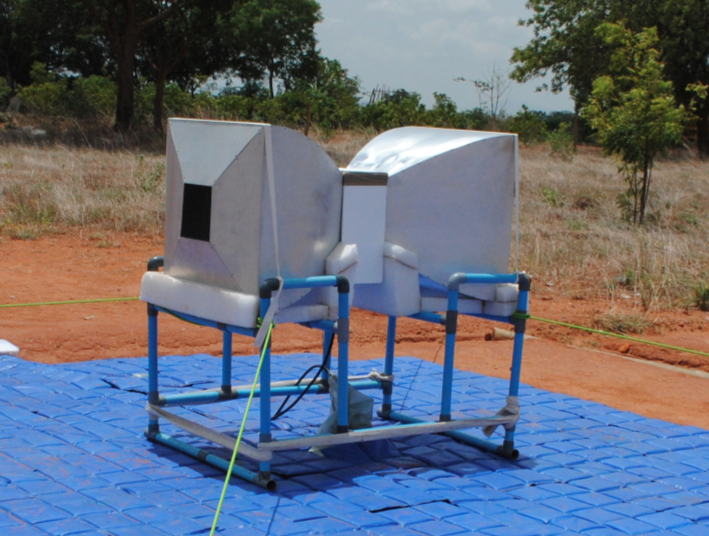}
\caption{ The SARAS Fat-dipole correlation spectrometer.}       
\label{fig:saras_photo}
\end{figure}

We have evolved a design for a correlation spectrometer for the measurement of the absolute spectrum of the cosmic radio background in the frequency range 87.5--175~MHz.  The system has been given the acronym SARAS, for Shaped Antenna measurement of the background RAdio Spectrum.  SARAS deployed in the Gauribidanur Observatory about 80~km north of Bangalore in India is shown in Fig.~\ref{fig:saras_photo}. In this paper we present the design of SARAS, with a focus on features---design strategies and calibration methods---that are novel and relevant for wideband measurements of background spectra.  The work presented herein represents progress towards building a system capable of a useful measurement of spectral signatures of the EoR in the cosmic radio background. 

We discuss the SARAS system configuration in Section~2 including details of the antenna as well as the analog and digital receivers. Because the work presented herein is a pioneering effort in that it presents a complex correlation approach to solving for total power spectra, we have chosen to present the performance in stages beginning with a simplified treatment and then going on to develop the understanding considering non-ideal behavior of increasing complexity. In Section~3 the measurement equations, signal processing and calibration strategies are first described assuming an ideal system performance.  Modified measurement equations taking into consideration first-order non-idealities are discussed in Section~4.1, with a more complete treatment considering  higher order corrections presented subsequently in Section~4.2.  The approach to modeling of SARAS data is discussed in Section~5.

\section{SARAS system configuration}
\label{sec:1}

We present here a system level design for a correlation spectrometer based receiver for the measurement of the absolute spectrum of the sky brightness.  The signal from the antenna is split early on, amplified by identical receiver chains, and processed by a digital spectrometer.  The two signals are sampled, the sampled data are discrete Fourier transformed, and data in corresponding frequency channels multiplied and the complex correlations integrated to form cross-spectra over the observing band.  

The primary motivation for adopting a correlation receiver configuration is that it makes possible switching to cancel spurious spectral features arising from additive correlator-based errors.  Digital receivers, which include digital signal processing devices that perform complex algorithms involve signals that are clocked over a range of rates and generate a broad spectrum of unwanted emissions that propagate across transmission paths via radiative and conductive coupling.  The high-speed clocking of digital signals also generates noise in system electronics ground.  Common mode coupling of these signals into the analog signal paths leading to the samplers results in spurious correlations, which may have frequency structure, and need to be canceled.  The system described here switches the antenna signal so that the sky signal appears in the complex spectrum with a sign that flips on alternate switch positions, where as the spurious additive correlations would be expected to remain unchanged, making it possible to subtract every set of spectra acquired in adjacent time intervals and in alternate switch positions and thereby cancel the spurious correlations.  

In a correlation receiver the antenna signal may be split prior to encountering any amplifier and, therefore, the system response does not contain the receiver noise temperature.  This source of noise in the signal path only appears in the measurement response of a correlation receiver because of unwanted leakage (non ideal performance) in the power splitter.  However, in the SARAS design, the switching that cancels additive correlator-based errors also cancels such leakage in the signal path.  

When calibrated for the complex bandpass, the sky power arriving at the antenna would appear wholly in the real component of the complex spectra measured by the correlation spectrometer.  However, reflections internal to the receiver chain cause signals to arrive at the samplers via multi-path propagation.  As discussed below, this is a route whereby receiver noise does appear in the measurement response.  The multi-path propagation and asymmetries in the system result in the measurement manifesting complex correlations in which the real and imaginary components share many descriptive parameters.  This provides an opportunity for using the imaginary components of the measured spectra, in which no sky signals are present, to estimate parameter values that characterize the multi-path related spectral features that contaminate the real spectra.  These parameters, which are derived from the measurements themselves,  may then be used in a joint modeling of (a) the components that constitute the system temperature, and (b) the propagation paths internal to the receiver, and thereby reveal the characteristics of the sky power arriving at the antenna.  This is a secondary advantage of adopting a correlation receiver configuration that includes a switch, and provides additional methods for modeling the sky power and distinguishing features owing to the cosmic radio background from multiplicative and additive features associated with internal reflections within the signal path.

\begin{figure}
\centering
\includegraphics[width=11cm]{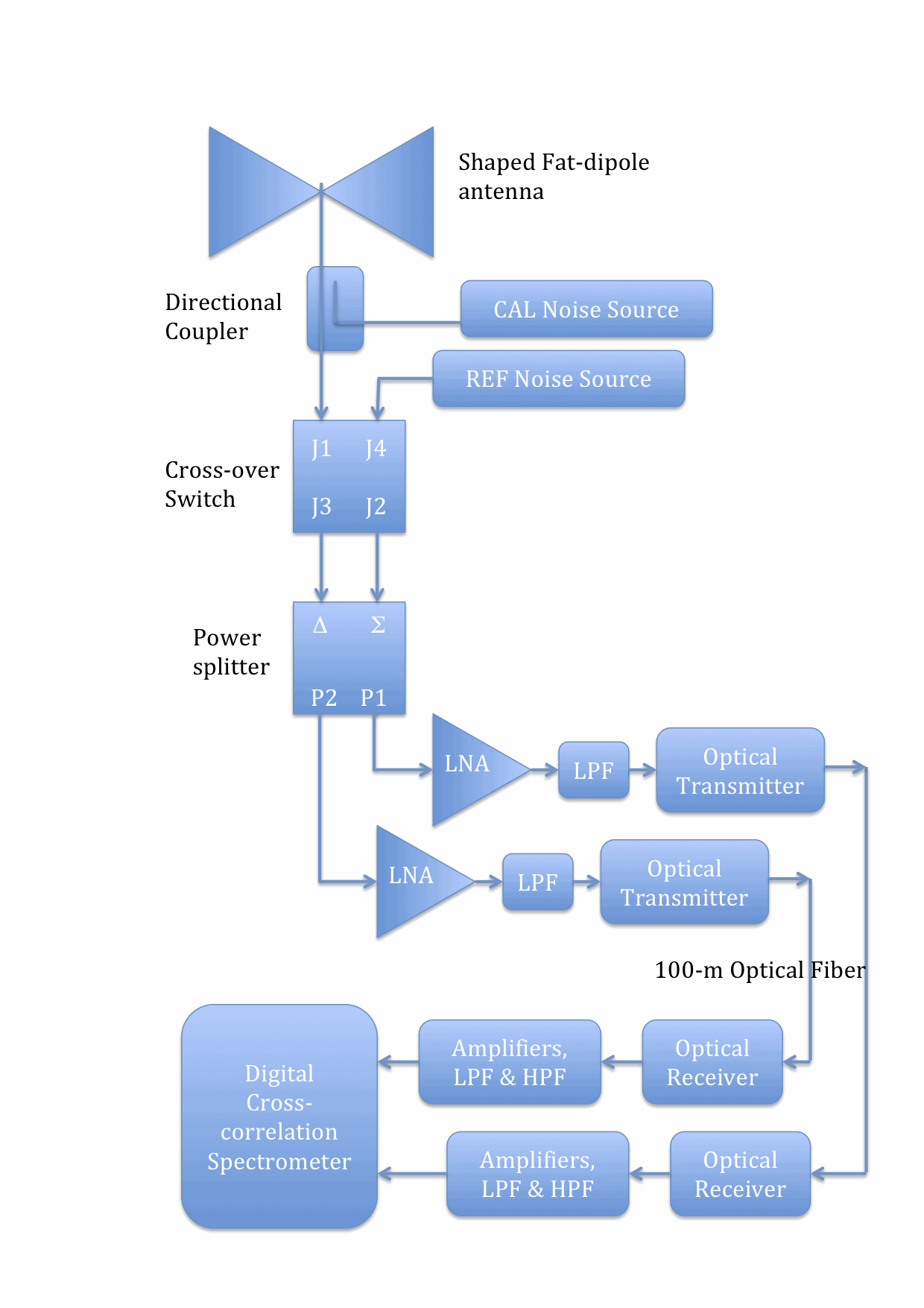}
\caption{The configuration of the SARAS spectrometer.}
\label{fig:design0}       
\end{figure}

The configuration of the SARAS spectrometer is shown in Fig.~\ref{fig:design0}.  In the sections below we describe each of the components in this signal path  and discuss the rational in the design.

\subsection{The Antenna: a frequency independent fat-dipole}

The antenna samples the sky brightness and couples the sky power to the signal path.  The sampling is defined by the response to the distribution of sky brightness and by the frequency dependence of this response.  The coupling efficiency of an antenna is related to the ohmic loss and the impedance match that the antenna offers to the electromagnetic wave arriving in free space.  The sky brightness has structure on a range of scales and frequency dependence in the antenna pattern will result in a spectral response that varies with frequency, which will depend on the frequency dependence of the antenna and the variation in sky brightness in the direction of the antenna beam.  For this reason a frequency independent antenna is desirable.  

An antenna that has physical size large compared to the wavelength of operation is also undesirable since any departures from frequency independent performance owing to manufacturing tolerances will result in spurious higher order components in the spectrum (owing to brightness structure on the sky) that may confuse cosmological signatures.  Antenna structures that are electrically long---like a frequency independent spiral antenna, for example---would have return loss and, therefore, a transfer function that would have high order structure.   High order frequency structure in the antenna transfer function will lead to spurious confusing spectral structure if there are errors in the calibration of the antenna response; therefore, it is desirable that the antenna return loss is smooth so that the system is tolerant to antenna calibration errors.  The above considerations require that the antenna be frequency independent, of small overall electrical size, and that the conductive parts be electrically small.  For the same reasons, it is desirable that any associated balun also be electrically small.

We use a frequency independent antenna that is electrically small: the antenna adopted for the SARAS spectral radiometer is a variant of the profiled fat-dipole (Raghunathan et al. 2012) designed to operate in the 87.5--175~MHz band.  The dipole elements have sine profiles and the antenna has been measured to have a beam pattern that deviates by less than 3\% from a cosine-squared radiation pattern, which is expected for a short dipole,  over the octave bandwidth.  The arms are of square cross section and constructed using aluminum sheets.  In place of the coaxial-choke balun adopted by Raghunathan et al. we have deployed a balun consisting of a transformer and a 180$^{\circ}$ hybrid, which are placed in between the arms and this module serves to transform the balanced out-of-phase signals from the two arms to an unbalanced signal on a coaxial cable.  This variation is motivated by the necessity to maintain short transmission lines between the antenna and receiver.

The antenna has a styrofoam base for structural stability and is placed on ground covered by Panashield SFA-type flat absorber tiles.  The height of the center of the dipole is 0.81~m from the tiles, which are of Nickel-Zinc Ferrite composition and have a reflectivity lower than $-30$~dB over the band.   The absorber ground affects the antenna temperature in two ways: (a) there is multi-path propagation of sky radiation to the antenna and from every sky direction the direct ray adds with a ray that arrives at the antenna after reflection from the absorber ground, the vector sum results in either constructive or destructive interference depending on the relative path delay and phase change on reflection; (b) the absorber ground emits with a brightness that depends on the ambient temperature and the reflectivity. Both these are indeed frequency dependent because the reflectivity is frequency dependent.   Our analysis of the consequent limitations to performance shows that the emission from the ground is an additive contribution to the antenna temperature that is expected to have a broad variation over the octave band with magnitude about 150~mK at the ends and significantly smaller at the band center, where the reflectivity is a minimum.  The multi-path propagation is of greater consequence and results in at most 1\% distortion to the sky spectrum.  

The ionosphere is a potential cause of spectral errors because refraction in the ionosphere shifts the apparent position of sources in the sky and, more importantly for SARAS, the shift would be dependent on frequency.  The effect of the ionosphere is minimized if the antenna pattern has low directivity and has a simple pattern without sidelobes.  This motivates our choice of a short dipole antenna, which has an omnidirectional pattern.

The small dependence of the antenna pattern on frequency over the octave operating band, the electrically short distance between the antenna and ground plane, along with the small reflectivity of the absorber tiles covering the ground, are together primarily a frequency dependent multiplicative gain that is subsumed into the calibration of the antenna gain: calibration is discussed below in Section~5.

\subsection{Analog receiver configuration}

The analog signal processing is in part implemented in a set of modules in a receiver unit at the antenna base and the remainder is in a base station about 100~m away where the digital receiver is also located.  This separation is necessary to prevent any self-generated RFI from the digital system entering the signal path via the antenna.  

The separation of the antenna base electronics from the digital receiver and the analog transmission of signals across this distance could result in multi-path propagation of signals owing to reflections.  Multi-path propagation that is confined to within each of the arms of the correlation receiver would effectively be a modulation of the individual band pass responses of the two arms, which would be expected to be calibrated as part of the complex bandpass calibration.  However, multi-path propagation that couples signals from one arm to the other would result in complex correlations that would leave residual modulations in the real and imaginary components of calibrated spectra.  Moreover, if signals propagate along the 100-m transmission line in a reverse direction from the base station to the antenna and if they then couple across to the second arm of the correlation receiver,  the spurious modulation in the calibrated spectra could have fine spectral structure with a period of about 1~MHz (depending on the speed of signal transmission in the 100-m line).  These unwanted couplings between the arms that occur at the antenna electronics  may arise because of reflections associated with impedance mismatch at the antenna terminals.  Alternately, unwanted coupling may also arise as a result of radiative leakage of signal power from the transmission line that is picked up by the antenna.  We have avoided these issues by implementing transmission across the 100~m distance on a pair of optical fibers, which provides excellent reverse isolation and also eliminates the potential for radiative couplings between the transmission lines and to the antenna.

As shown in Fig.~\ref{fig:design0}, the signal from the antenna and a calibrated broadband noise source (referred to as CAL) are fed to the direct and coupled port respectively of a directional coupler. The broadband noise source provides an excess calibration noise when it is `ON'.  The output of the directional coupler, which consists of antenna noise plus calibration signal (when the CAL in ON) is fed to port J1 of a cross-over switch.  The second input port J4 of this switch is fed power from a second noise source that serves as a reference (referred to as REF).   As discussed below, the recorded spectra in the SARAS configuration are always difference spectra: what is measured is the difference between the powers from the antenna (plus CAL power when the calibration noise source is ON) and reference (the value of REF power depends on whether or not the reference noise source is on).  The reference is an ambient temperature noise source when the second REF noise source is in the OFF state, and provides an excess reference power when it is in the ON state. The difference between spectra recorded in the two states of the reference noise source is used to derive additional constraints while modeling the systematics. The output ports J2 and J3 of the cross-over switch are connected respectively to the sum and difference ports of a 4-port power splitter, whose output ports $P_{1}$ and $P_{2}$ are connected to independent low-noise amplifiers at the head of a pair of separate analog receiver chains.  The cross-over switch is electro-mechanical, with low loss and high isolation.  At the antenna base, the pair of analog chains include a low-noise amplifier and a low pass filter, which are followed by modular fiber-optic transmitters that include transconductance amplifiers and couple the electrical signal into single-mode 1550~nm optical fibers.  

The calibration noise source is a 25~dB excess noise source followed by a 3~dB attenuator that connects the calibration noise to the coupled port of the directional coupler.  The reference load consists of another 25~dB excess noise source followed immediately by a 23~dB attenuator.   In both cases the equivalent temperatures added by the calibration noise and reference noise are similar to the system temperature when both noise sources are off.

At the base station, the analog receiver includes an optical receiver followed by a cascade of low and high pass filters with amplifiers and attenuators in between.  The filters limit the band between 87.5 to 175 MHz; for initial field testing in sites where FM transmission is a potential source of RFI, high pass filters limit the lower end of the band to 110~MHz.

\subsection{Digital receiver configuration}

The analog signals are sampled at 175~MHz.  The sampled data streams are discrete Fourier transformed to decompose the signals over 1024 frequency channels.  Complex data streams from the two receiver chains at corresponding frequencies are multiplied and the complex products accumulated to represent cross-power spectra.  Autocorrelation spectra of the digital signals in the two receiver chains are also computed.  These are all read into a laptop PC and stored for offline processing.  The digital signal processing is performed on a single Virtex-5 FPGA chip.

The digital receiver includes a GPS disciplined frequency reference that provides sufficient long term stability, a frequency synthesizer that generates the sampling clock and a digital i/o card for generating signals that switch the calibration noise, reference noise and cross-over switch.  All switching signals are sent to the antenna base via optically isolated copper transmission lines.

The digital systems are in a shielded enclosure that is sealed from the environment.   Cooling is achieved using heat pipes which exchange heat within the enclosure and to external air by forced convection using fans.  The d.c. fans are brushless; however, in order to drive the fan rotor its position is sensed by a Hall-effect sensor and the stator fields are flipped by switching the current direction.   This produces measurable broad band RFI from the digital receiver because a pair of fans are necessarily mounted outside the sealed enclosure: a noise power of order 10~K enters the antenna if such fans are located 1~m from the antenna without shielding.  We have suppressed this contribution to be less than 1~mK by including RFI shielding in the air vents of the fan enclosure and locating the digital receiver 100-m from the antenna.
         
 \section{Signal processing in the SARAS radiometer}
 
\begin{figure}
\centering
\includegraphics[width=13cm]{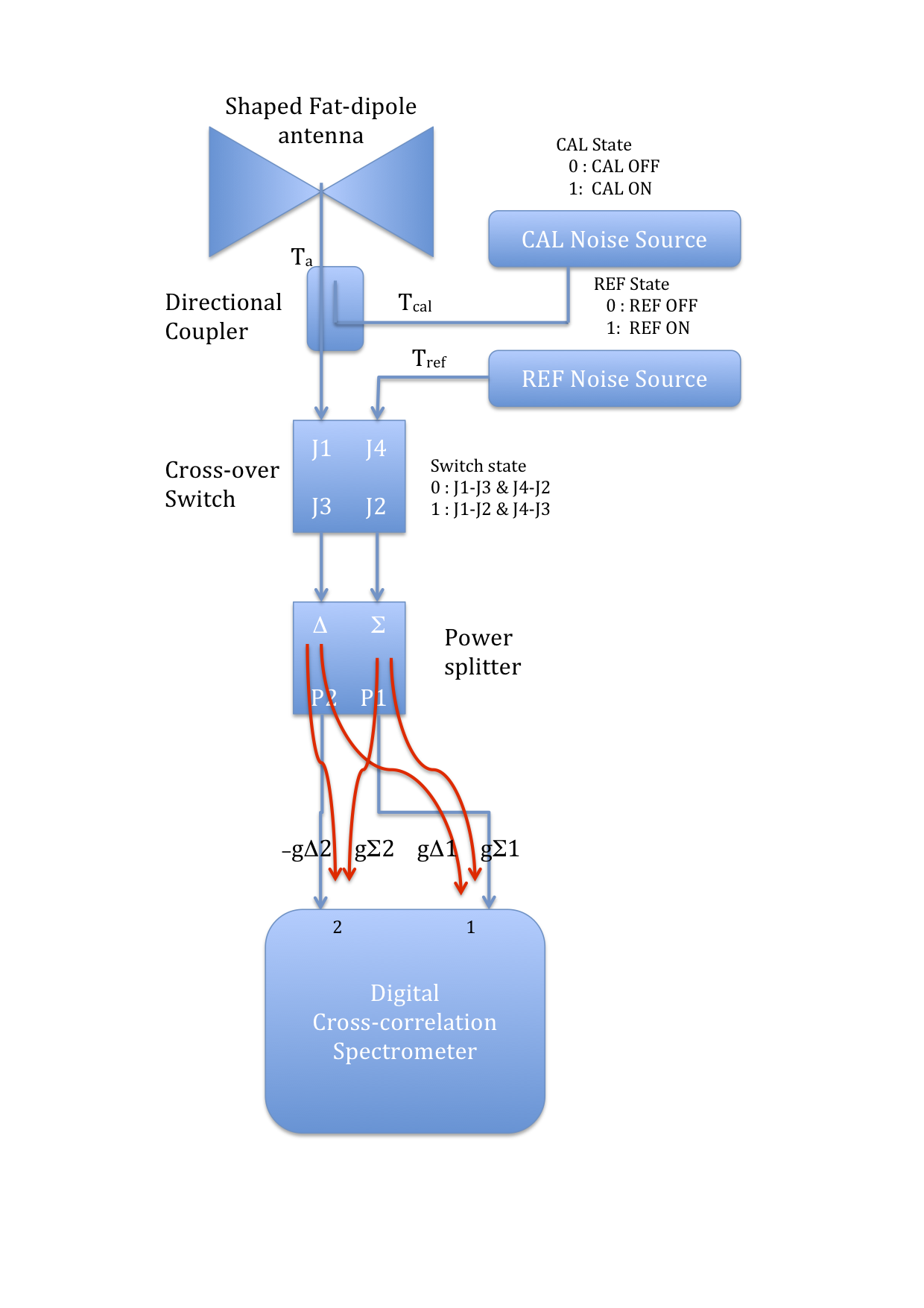}
\caption{Depicts the states of the switch and noise sources and shows the gains in the signal paths.}
\label{fig:design1}       
\end{figure}

Let $T_{a}$ represent the antenna temperature at the mainline input port of the directional coupler, $T_{cal}$ represents the excess calibration noise (CAL) at the coupled port of the directional coupler when the CAL is switched on, $T_{ref0}$ represents the reference temperature at the second port of the cross-over switch when the reference noise source (REF) is in `off' state and $T_{ref1}$ is the excess noise when REF is in `on' state.  All temperatures are referred to the mainline input port of the directional coupler, where the antenna would be connected.  $g_{\Sigma 1}$ and $g_{\Sigma 2}$ represent the gains of the paths from the $\Sigma$ port of the power splitter to the two samplers in the digital cross-correlation spectrometer, $g_{\Delta 1}$ and $g_{\Delta 2}$ are the gains of the paths from the $\Delta$ port to the two samplers. All of the temperatures and gains are functions of frequency; the $g$-terms represent voltage gains.  For clarity, we herein choose to represent the voltage gains between the $\Delta$ and $\Sigma$ ports and $P1$ and $P2$ ports of the power splitter by terms that are ideally unity, and explicitly include a negative sign for the voltage transfer function between $\Delta$ and $P2$.  These symbols, the conventions we adopt for denoting switch positions and states of the noise sources are depicted in Fig.~\ref{fig:design1}.  

The measurement and calibration sequence consists of alternately switching the mainline output of the directional coupler and the reference noise source between the $\Sigma$ and $\Delta$ ports of the power splitter by means of the cross-over switch.  At each of the positions of the cross-over switch, the calibration noise CAL is switched on for a time $\tau_{cal}$  while the reference noise source is off, the reference noise REF is switched on for a time $\tau_{ref}$ while the calibration noise is off, and for a time $\tau_{src}$ both sources of noise are off.    We denote the `direct' switch position with the antenna connected to $\Delta$ port as `0', and the `crossed' switch position with the antenna connected to $\Sigma$ port as `1'.  Complex spectra are acquired with the system cycling through a set of six states listed sequentially in the successive rows of Table~\ref{table:states}.  $\tau_{cal}$, $\tau_{src}$ and $\tau_{ref}$ are all nominally the same time and the integration time is 0.7~s in each state.  

\begin{table}
\caption{The observing schedule with SARAS. The system cycles through the states listed in the rows of the table and time averaged complex spectra are recorded separately and sequentially corresponding to each of these states.  }
\label{table:states}
\centering
\begin{tabular}{ccc}
\hline\noalign{\smallskip}
Switch state & Cal state & Ref state  \\
\noalign{\smallskip}\hline\noalign{\smallskip}
0 & 0 & 0 \\
0 & 1 & 0 \\
0 & 0 & 1 \\
1 & 0 & 0 \\
1 & 1 & 0 \\
1 & 0 & 1 \\
\noalign{\smallskip}\hline
\end{tabular}
\end{table}

In the remainder of this section, we present an analysis of the signal processing and calibration assuming ideal impedance matching between components and transmission lines. In later sections we deal with the issues related to impedance mismatches.

\subsection{Measurement equations}

The antenna signal (or the antenna signal plus the calibration noise) and reference noise are connected to the sum ($\Sigma$) and difference ($\Delta$) ports of the power splitter.  Therefore, the cross-power spectrum measured by SARAS is at all times the difference between the antenna temperature and reference.  The receiver noise does not appear in the response because (a) the noise in the receivers in the two signal paths following the power splitter are uncorrelated and (b) under the assumption that the interfaces are impedance matched and the antenna has zero return loss the receiver noise from one arm of the correlation receiver does not leak into the other arm. 

The measured complex cross powers in the two switch states when both the reference and calibration noise sources are off may be written as:
\begin{equation} 
P_{0off} = -g_{\Delta 1}g_{\Delta 2} T_{a} +g_{\Sigma 1}g_{\Sigma 2}T_{ref0} + P_{cor}   
\end{equation}
and
\begin{equation}                    
P_{1off} = g_{\Sigma 1}g_{\Sigma 2} T_{a}  - g_{\Delta 1}g_{\Delta 2} T_{ref0} +P_{cor} .              
\end{equation}
$P_{cor}$ denotes the spurious additive complex cross-power spectrum that may be present in the measured spectrum because of undesired coupling of power between the transmission paths and samplers in the two arms.  $P_{cor}$ is expected to be a constant and independent of the state of the cross-over switch and the state of the calibration and reference noise sources.  The measured complex cross powers in the two switch states when the calibration noise alone is on may be written as:
\begin{equation}
             P_{0cal} = -g_{\Delta 1}g_{\Delta 2} (T_{a}+T_{cal}) 
                                +g_{\Sigma 1}g_{\Sigma 2} T_{ref0}
                                +P_{cor}
\end{equation}
and
\begin{equation}
               P_{1cal} = g_{\Sigma 1}g_{\Sigma 2} (T_{a}+T_{cal}) 
                              -g_{\Delta 1}g_{\Delta 2} T_{ref0}
                              +P_{cor}.
\end{equation}
The measured complex cross powers when the reference noise source alone is on may be written as:
\begin{equation}                    
               P_{0ref} = -g_{\Delta 1}g_{\Delta 2} T_{a}
                                +g_{\Sigma 1}g_{\Sigma 2} T_{ref1}
                                +P_{cor}      
\end{equation}
and
\begin{equation}
               P_{1ref} = g_{\Sigma 1}g_{\Sigma 2} T_{a}
                              -g_{\Delta 1}g_{\Delta 2} T_{ref1} 
                              +P_{cor}.
\end{equation}
The subscripts `0' and `1' for the powers denote spectra for the two positions of the cross-over switch.   All of the above powers are functions of frequency.  

\subsection{Cancellation of spurious additive component of the response}

In the two switch positions the antenna (and the calibration signal) and reference are connected to the $\Sigma$ and $\Delta$ ports of the power splitter alternately. Therefore, in the two switch positions, the difference spectra---difference between the signals at the $\Sigma$ and $\Delta$ ports of the power splitter---appear in the measurement with opposite sign.  However, in both switch positions the magnitude and sign of the spurious additive contribution $P_{cor}$  that arises because of unwanted couplings of noise across the two samplers or between the pair of signal paths from the power splitter to the samplers remains unchanged. Differencing the `0' and `1' spectra corresponding to the two switch positions, therefore, cancels any such spurious additive contribution to the measured spectra. 

The difference cross-power spectra are     
\begin{eqnarray}
P_{off} & = & P_{1off}-P_{0off} \nonumber \\
 & = &  ({g_{\Sigma1}g_{\Sigma2}+g_{\Delta1}g_{\Delta2}}) (T_{a} - T_{ref0}),
\end{eqnarray}
\begin{eqnarray}
P_{cal} & = & P_{1cal}-P_{0cal} \nonumber  \\
 & = & ({g_{\Sigma1}g_{\Sigma2}+g_{\Delta1}g_{\Delta2}}) (T_{a}+T_{cal} - T_{ref0})
\end{eqnarray}
and
\begin{eqnarray}
P_{ref} & = & P_{1ref}-P_{0ref} \nonumber \\
 & = &  ({g_{\Sigma1}g_{\Sigma2}+g_{\Delta1}g_{\Delta2}}) (T_{a} - T_{ref1}).
\end{eqnarray}

 \subsection{Calibration for the complex instrument bandpass}
 
 The calibration noise source is coupled to the signal path from the antenna immediately following the balun.  Along with the antenna signal, the calibration noise is also connected alternately to the $\Sigma$ and $\Delta$ ports of the power splitter, and in each switch position the complex cross-correlation spectrum is recorded separately with the noise source switched on and off.  Neglecting the asymmetries in the switch and differences in the reflection and transmission properties of the different switch paths, the calibration noise may be used to calibrate the antenna temperature for the bandpass of the signal path downstream of the cross-over switch.  We assume here that the excess noise at the reference port of the cross-over switch has a flat temperature spectrum. 

Subtraction of the $P_{off}$ spectrum from the $P_{cal}$ spectrum gives the instrument response to the calibration noise:
\begin{equation}
     P_{cal}-P_{off} = (g_{\Sigma1}g_{\Sigma2}+g_{\Delta1}g_{\Delta2}) T_{cal}.
\end{equation}
Since the calibration noise samples both the $\Sigma$ and $\Delta$ ports of the power splitter, the above power spectrum that represents the excess calibration noise includes the mean multiplicative gains of both ports.

The difference spectrum $P_{off}$  is bandpass calibrated using $(P_{cal}-P_{off})$ and scaled 
by the calibration noise temperature $T_{cal}$ to derive a calibrated spectrum in antenna temperature units.  This calibration yields the desired antenna temperature minus the temperature of the reference port:
\begin{equation}
        T_{a}  - T_{ref0} =  {P_{off}\over(P_{cal}-P_{off})} \times T_{cal}.               
\end{equation}
It may be noted here that the gains $g_{\Sigma 1}$, $g_{\Sigma 2}$, $g_{\Delta 1}$, and $g_{\Delta 2}$ cancel in this process.  It is to make this happen that the SARAS configuration is designed so that the calibration noise is switched on and off in both switch states and the cross-over switch directs both antenna noise and calibration noise temperatures via both the $\Sigma$ and $\Delta$ ports of the power splitter.

Following calibration for the complex bandpass, the real component of the calibrated cross-spectrum represents the antenna temperature minus noise temperature of the reference.  At frequencies where the antenna temperature exceeds the reference temperature the real component will be positive, where as over frequencies where the antenna temperature is less than the reference noise temperature the real component will be negative.  The imaginary component is expected to vanish in the ideal case considered in this section.

If the impedance match along the signal paths were excellent (ideal) then the reference noise is unnecessary.  However, as discussed below, this second noise source has been included in the SARAS configuration to serve as a method of deriving a template for the instrument response to the reference load.  The SARAS measurement of the sky spectrum is a differential measurement in that it is the difference between antenna temperature and that of the reference load; the calibration noise serves to provide a template for the instrument response to the antenna signal where as the reference noise source serves to provide a template for the instrument response to the reference load.  It may also be mentioned here that an additional source of significant noise temperature in the receiver is the noise temperature of the first low-noise amplifiers in the two arms of the correlation spectrometer.  Although in an ideal system this noise would not contribute to the measured spectrum; however, impedance mismatches and leakage terms in the power splitter result in a response to receiver noise and it would be useful to have low-noise amplifiers whose noise temperatures may be switched between nominal and high states!  We are not aware of how such a feature may be implemented in practice.
      
If we assume that the calibration noise temperature is a constant over the octave observing band, then $T_{cal}$ is independent of frequency.  The value of $T_{cal}$ is determined by replacing the antenna with terminations at known temperatures and comparing the step in measured power with the step measured when the calibration noise is switched between on and off states.

\subsection{Calibration for the antenna gain}

If the antenna were ideal in that the antenna impedance matched free space and the ohmic losses are negligible, the antenna temperature would be the mean brightness temperature over the radiation pattern, where the weighting is the normalized radiation pattern.  The frequency dependance of the gain loss due to impedance mismatch and ohmic loss may be calibrated by modeling the difference spectra when the sky brightness in the main lobe of the beam changes significantly, assuming that we know the difference spectrum of the sky brightness.  We examine here considerations for the modeling of the sky brightness that are needed for calibration of the antenna gain.

The cosmic radio background  towards most sky directions are very nearly power-law spectra, with a temperature index that has a mean of about $-2.55$ (Kogut et al. 2011; Seiffert et al. 2011).  The index varies somewhat with direction.  We adopt the 408-MHz sky survey of Haslam et al. (1982) as representing the sky brightness distribution, and assume that the spectral index towards any sky direction is given by the 2-point spectral index computed between this image and the 150-MHz all-sky map synthesized by Landecker and Wielebinski (1970).   The 408-MHz image has a quoted uncertainty of $\pm3$~K  in the zero point and $10\%$ in absolute scale; the 150-MHz image has uncertainties of $\pm20$~K  in the zero point and $4\%$ in absolute scale.  For this assumed sky model, the brightness temperature of the sky at the center of the SARAS band is about $210 \pm 60$ ~K in the coldest regions and the brightness rises to several thousand Kelvin towards the Galactic Center.     

\begin{figure}
\centering
\includegraphics[width=6cm,angle=-90]{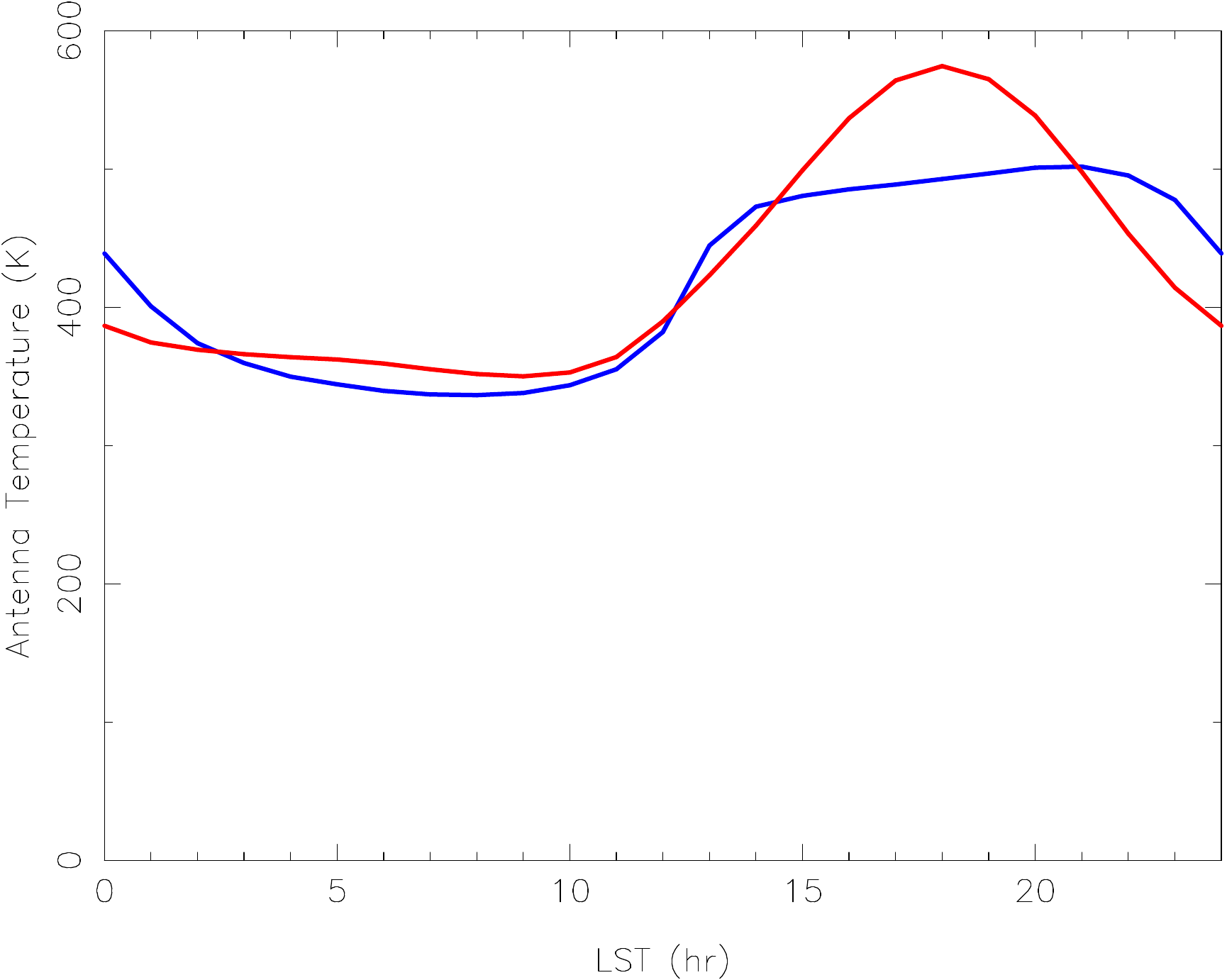}
\caption{Expected variation in antenna temperature over LST at the band center frequency of 131.25 MHz. The red curve is for EW dipole orientation and blue curve is for NS orientation.}
\label{fig:tsys_lst}       
\end{figure}

The SARAS fat-dipole antenna has a dipole radiation pattern that averages over a beam solid angle covering two-thirds of the sky.  Adopting the above model for the brightness distribution over the sky and across the frequency band, the SARAS dipole has an antenna temperature at the band center frequency of 131.25~MHz whose variation with LST is as given in Fig.~\ref{fig:tsys_lst}.  The ground has been assumed to be a perfect absorber with brightness 300~K, and the variation in antenna temperature is displayed for an EW as well as NS orientation for the dipole.  A location at $+13^{\circ}$ latitude has been assumed.  When the system is used to detect spectral signatures in cosmological radio backgrounds such as the all-sky signal corresponding to the epoch of reionization, a NS orientation is preferred since this orientation has a lower antenna temperature over a wider LST range, and hence the sensitivity to any spectral signatures would be greater.

\begin{figure}
\centering
\includegraphics[width=6cm,angle=-90]{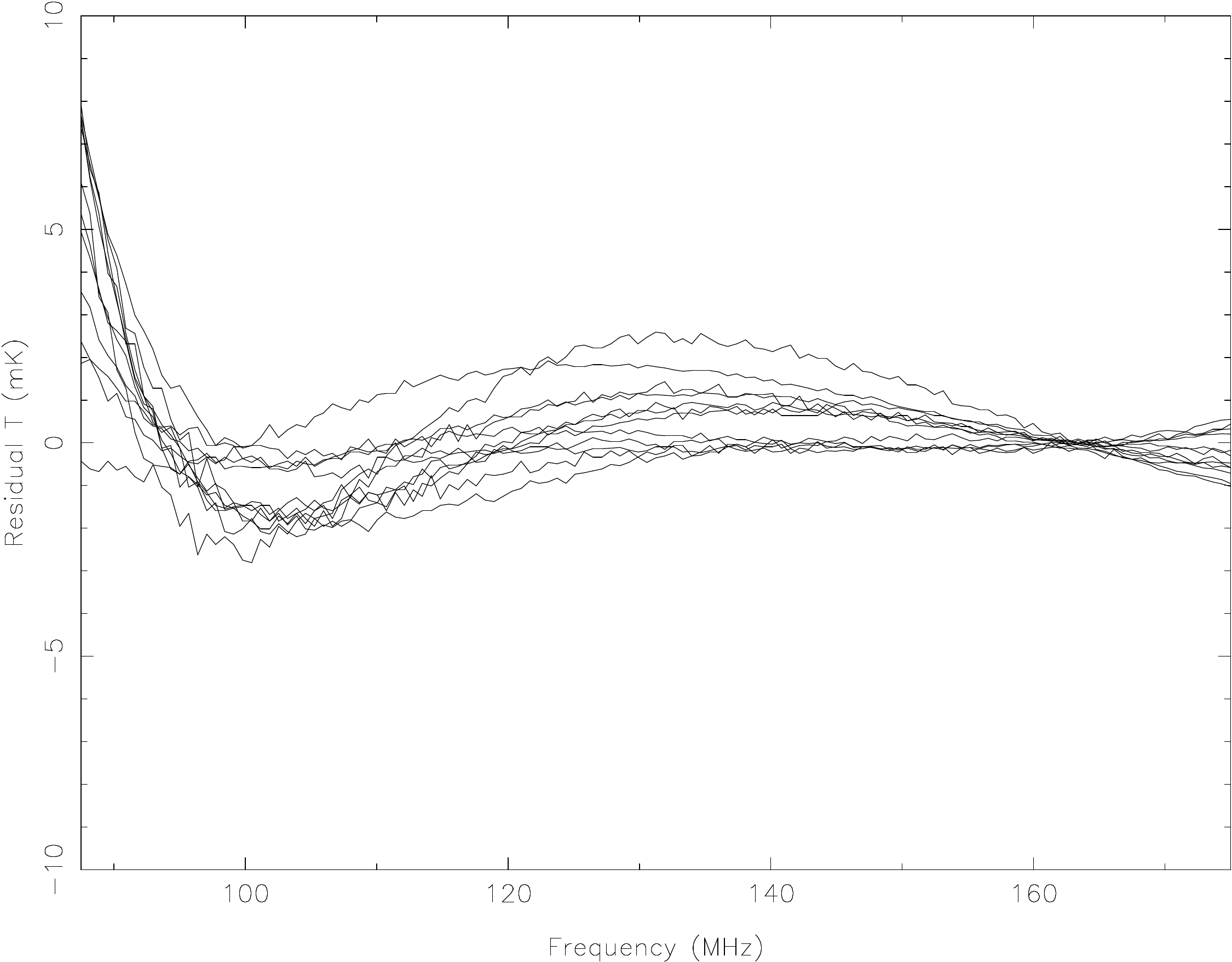}
\caption{Residuals to fits to synthetic sky spectra.  Third order polynomials of the form given in equation~9 were fitted to synthetic spectra.}
\label{fig:residT}       
\end{figure}

We assume that the spectra of the sky brightness towards any direction are of simple power-law form.  At any instant, the measured sky spectrum is a linear combination of the power-law form sky spectra towards directions within the beam.  This measured spectrum will no longer be a single power law and will require a more complex fitting form.  We have computed synthetic spectra that would be observed with a short dipole antenna over LST and attempted to fit the distribution of log(T) versus log(frequency) using polynomials.   The averaging over power-law spectra results in a spectrum that is, unsurprisingly, dominated at the highest frequencies by the sky regions within the beam that have relatively flatter spectra, and is dominated at the lowest frequencies by sky regions that have the steepest spectral indices.  Therefore, relative to the best fit power-law model, the residuals are concave upwards in shape.  We have attempted a fit of third order polynomials to the synthetic spectra; our fit takes the form:

\begin{equation}
{\rm log}_{10} T = a_{0} + a_{1} (log_{10}\nu )+ a_{2} (log_{10}\nu)^2 + a_{3} (log_{10}\nu)^3
\label{fg}
\end{equation}
where $T$ is the temperature and $\nu$ is the frequency.  In order that the fitting function be smooth and without inflection points, the coefficients $a_{2}$ and $a_{3}$ are constrained to be positive.  The residuals of this fit to spectra that were synthesized at intervals of 2~hr in LST are shown in Fig.~\ref{fig:residT}, and are within a few mK over most of the frequency range, suggesting that sky models of this form should suffice.  It may be noted here that Pritchard and Loeb (2010) arrived at similar conclusions based on their fits to synthetic sky spectra that were derived using the global sky model of de Oliveira-Costa et al. (2008).

The difference between sky spectra measured towards different directions is again essentially a linear combination of power laws corresponding to the Galactic emission, since it is expected that the uniform extragalactic component including any re-ionization signatures would be canceled in the sky difference.   The difference of spectra obtained during LST when the antenna temperature is relatively high and that when the antenna temperature is relatively low serves as a smooth calibration signal.  We have divided the synthetic spectra computed over LST into two halves based on the median antenna temperature, computed the difference spectrum, and examined the goodness of fit when fitting to the form described in Equation~12.  The residuals to this fit are within $\pm 1$~mK over most of the band.

To summarize the discussion in this section, it is sufficient and necessary that the sky temperature spectrum is modeled with four parameters and we may adopt the form given in Equation~12.  The coefficients $a_{2}$ and $a_{3}$ may be constrained to be positive, so that the model is smooth and does not contain inflections.

\section{SARAS radiometer: higher order corrections for non-ideal behavior}

The measurement equations and calibration method described above assumes ideal signal propagation paths that are impedance matched at all interconnections and that there are no reflections at the interfaces.  The antenna has also been assumed to be impedance matched to free space and to the transmission line.  Practical antennas, devices, as well as transmission lines have impedances that deviate from the nominal design value and this may vary with frequency resulting in frequency dependent reflection of signals at interfaces.  Additionally, when long transmission lines are involved, there may be impedance variations along the path owing to manufacturing tolerances and this may result in distributed reflections of the propagating signals.  In this section we discuss the modeling of departures from the ideal behavior analyzed earlier, which arise from impedance mismatches.

 The impedance mismatch at the antenna is characterized by its complex reflection coefficients---also referred 
 to as `return loss'---which is usually frequency dependent.  Hence only a part of the electromagnetic radiation incident on the antenna is coupled to the transmission line.   This signal loss is frequency dependent and is over and above the direction dependent response defined by the normalized radiation pattern as well as any ohmic loss, which may be independent of the direction of incidence.  
 
 In a real system, the LNA input impedance is complex and differs from the impedance of the components and transmission lines which interconnect the antenna and reference load to the amplifier inputs. The impedance mismatch at the LNA inputs are also characterized by complex reflection coefficients which may be functions of frequency.
 
 The dominant sources of noise power in the signal paths within the SARAS radiometer configuration are (a) the antenna temperature, (b) noise power from the reference load, (c)  noise from the first stage of low-noise amplifiers and, (d) when the calibration noise is on, the additional noise from that source.   Amplifier noise has a component that adds to the input signal and propagates downstream along with the amplified input signal; it has a second component---correlated with the first---that propagates upstream against the nominal signal flow direction.  All noise in the signal path that propagates upstream towards the antenna would be ideally expected to exit the system via the antenna; however, these are partly reflected by the mismatch at the antenna, which is characterized by the antenna reflection coefficient or return loss.

In the SARAS configuration, the electronics at the antenna base are all packaged with modules directly connected successively or with absolute minimum lead lengths.  Nevertheless, the antenna temperature, reverse propagating receiver noise from the two low-noise amplifiers and noise power from the reference load (and calibration noise) would undergo multiple reflections between the antenna, reference load and amplifier input terminals.  Reflections in this part of the radiometer between terminations on either side of the cross-over switch and power splitter give rise to spurious responses that are not canceled in the switching; the reflections at this stage couples power across the two arms of the correlation spectrometer causing the radiometer to respond to receiver noise as well.  Multi-path propagation leads to spurious complex responses that appear in the real and imaginary components of the measurements as ripples whose period in frequency space depends on the excess traversal delay in the multi-path propagation. Since the spurious correlations arising from internal reflections across the cross-over switch and power splitter appear as complex responses, where as the ideal system response is purely real, the SARAS system configuration allows a modeling of the spurious additives to the real spectrum by modeling the ripples in the imaginary spectrum.

\begin{figure}
%\begin{minipage}[b]{0.5\linewidth}
\centering
\includegraphics[ angle=270, width=8cm]{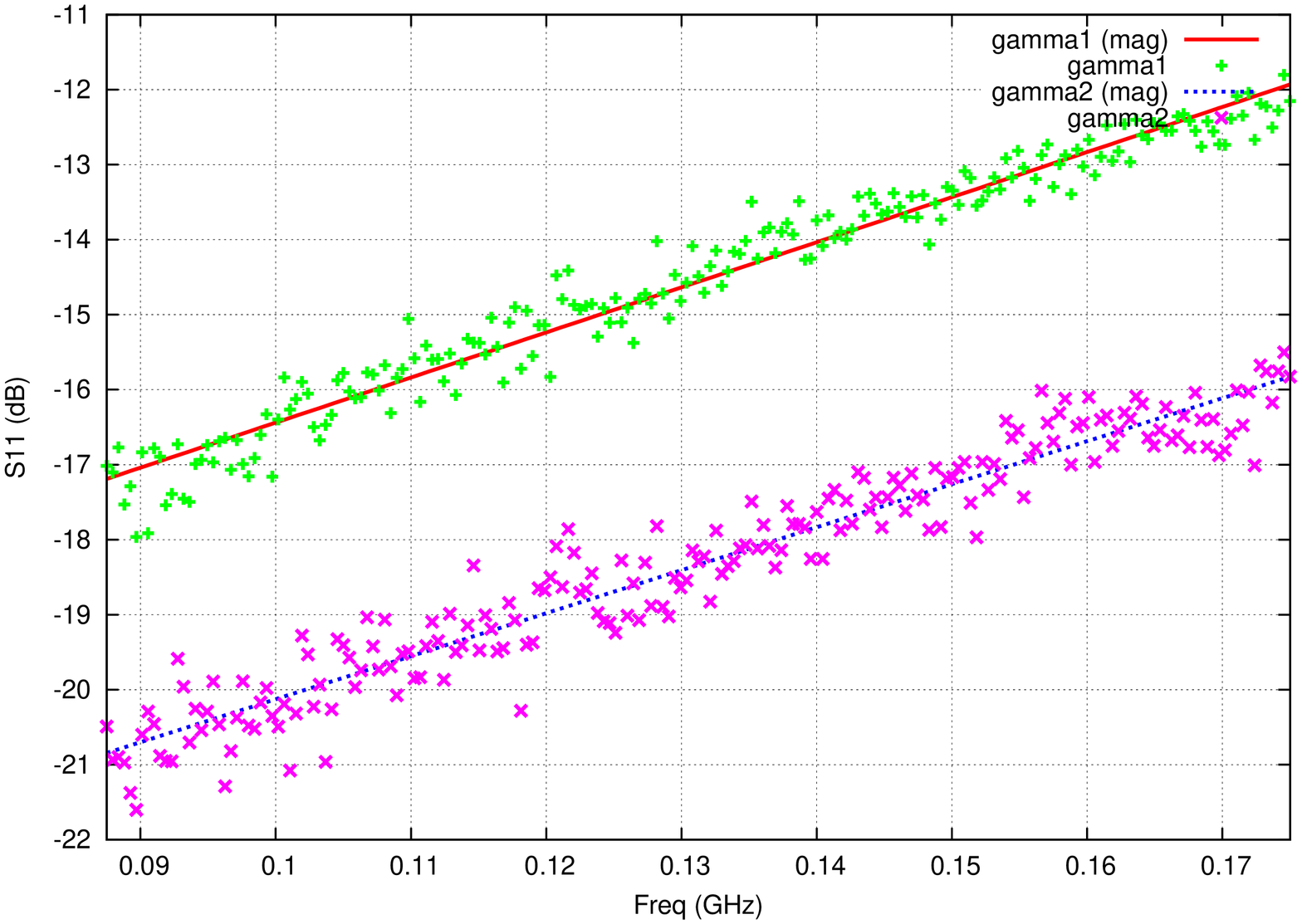}
\caption{Reflection coefficients of the two LNAs (Magnitude).}
\label{fig:refcofM}
%\hspace{0.5cm}
%\begin{minipage}[b]{0.5\linewidth}
\centering
\includegraphics[ angle=270, width=8cm]{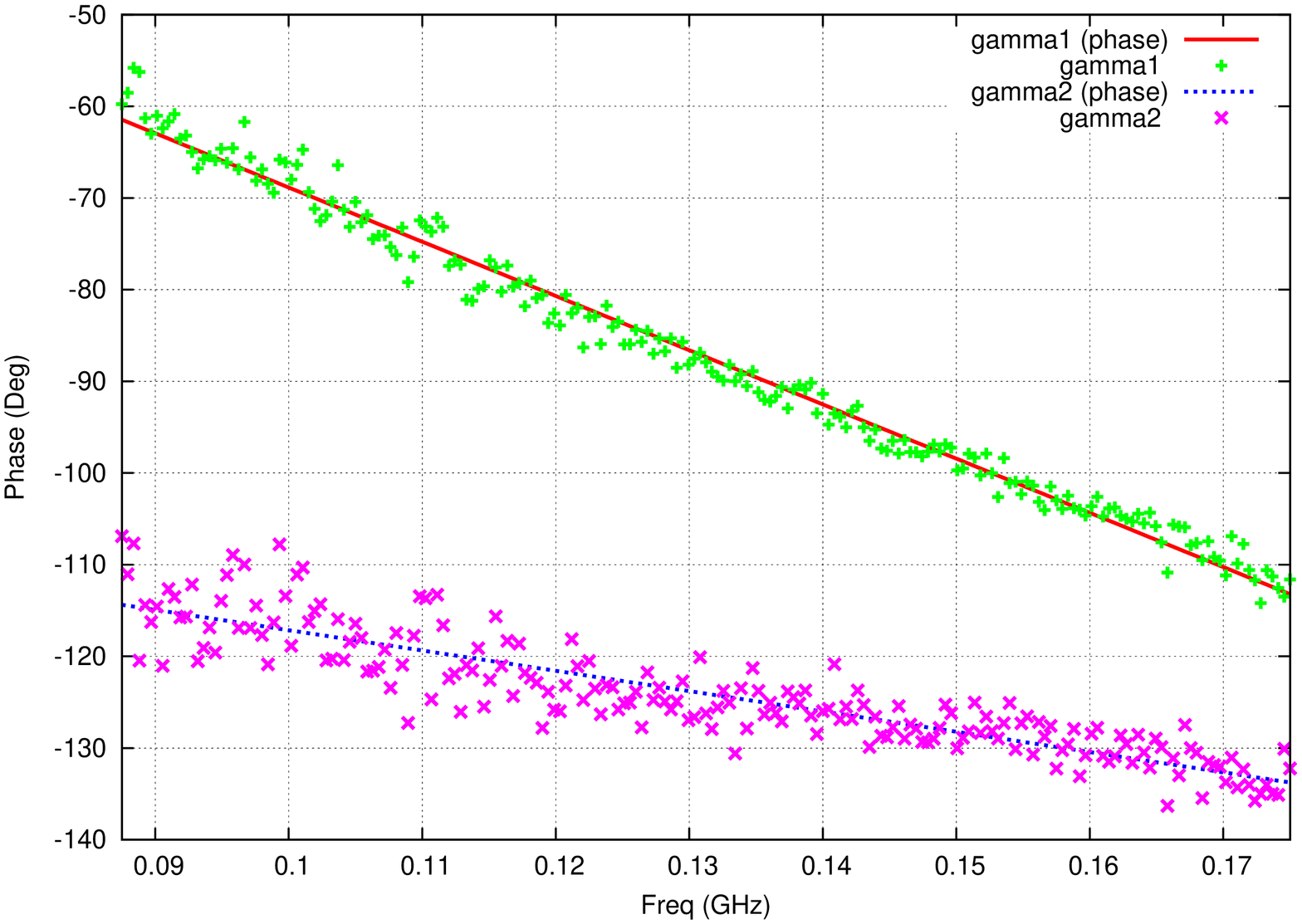}
\caption{Reflection coefficients  of the two LNAs (Phase).}
\label{fig:refcofp}
%\end{minipage}
\end{figure}

\begin{figure}
\centering
\includegraphics[ angle=270, width=8cm]{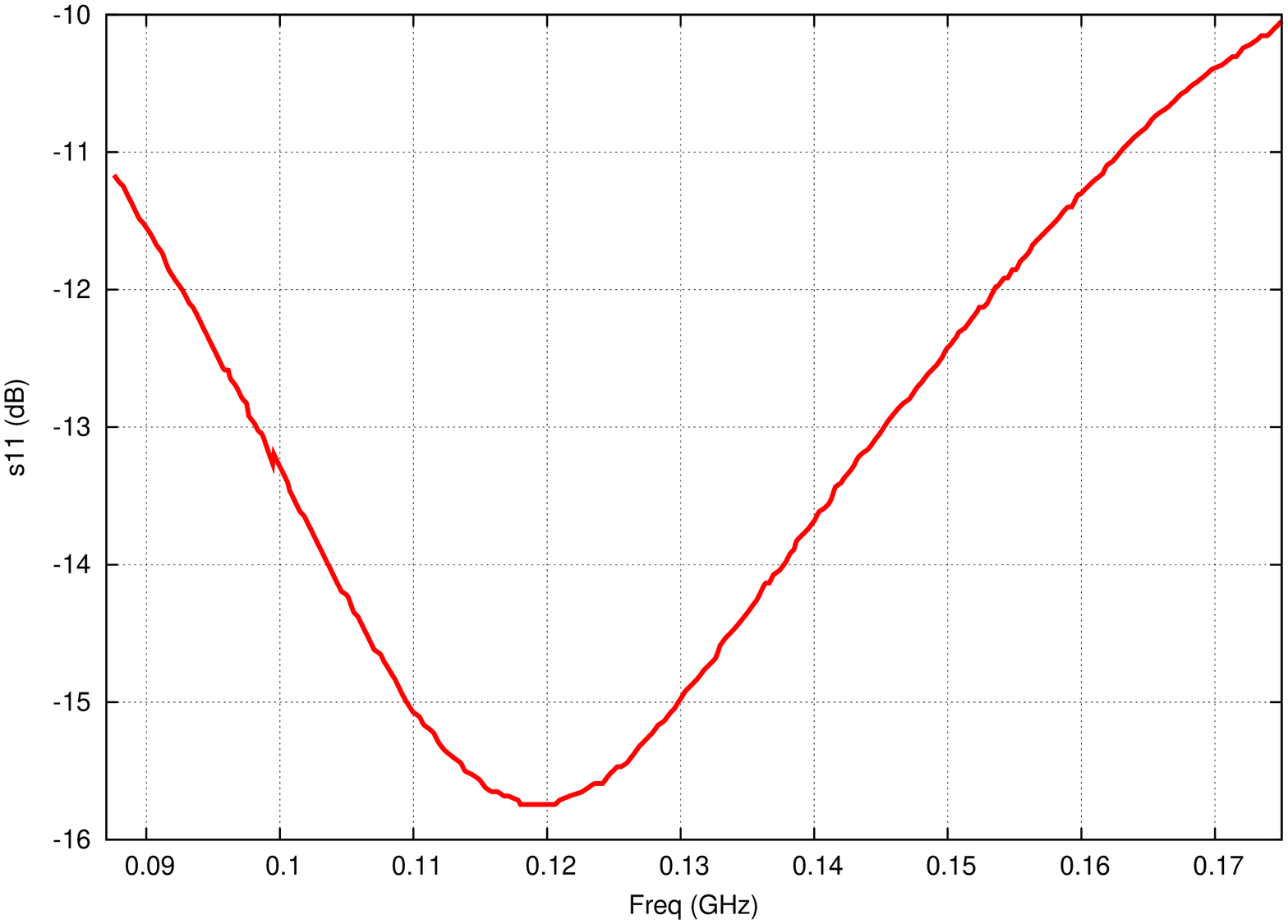}
\caption{Reflection coefficients of the antenna (Magnitude).}
\label{fig:ant_rlM}
\centering
\includegraphics[ angle=270, width=8cm]{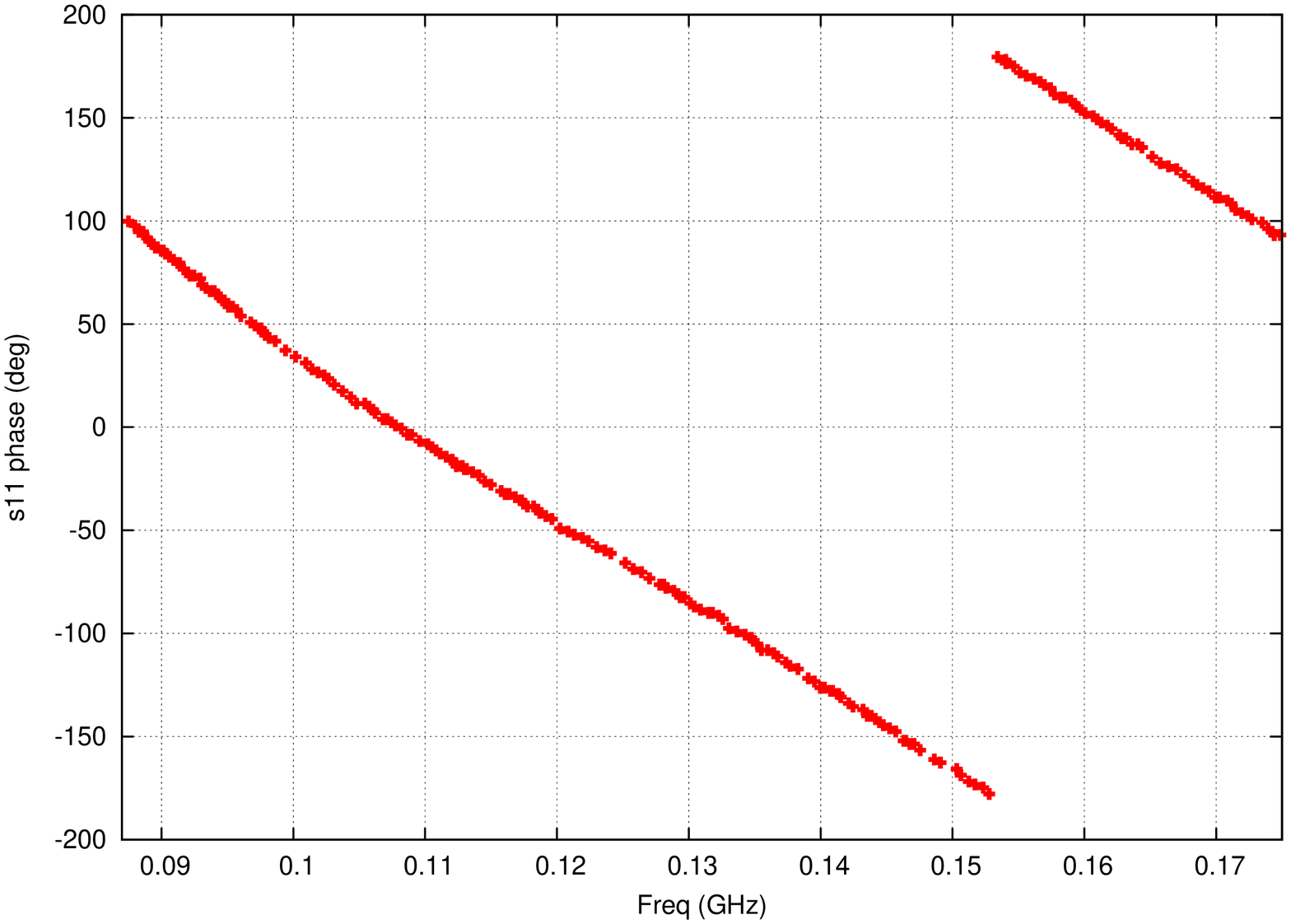}
\caption{Reflection coefficients  of the antenna (Phase).}
\label{fig:ant_rlP}
\end{figure}

We denote by symbols $\Gamma_a$, $\Gamma_{ref} $, $\Gamma_1$ and $\Gamma_2$  the complex reflection coefficients at the antenna, reference  and at the inputs of the two receivers respectively. In Figs.~\ref{fig:refcofM} \& \ref{fig:refcofp} we show the measured complex reflection coefficients at the inputs of the SARAS low-noise amplifiers in the form of plots of the amplitude of the reflected power, and phase of the reflected voltage, as functions of frequency.  The amplifiers we have used are not a matched pair and, therefore, the reflectivities differ.  Additionally, these functions appear smooth over frequency because the amplifiers are wide-band and have operating frequency range much wider than the SARAS system.  In Figs.~\ref{fig:ant_rlM} \& \ref{fig:ant_rlP} we show corresponding plots for the measured reflection coefficient at the antenna.   

The measured complex reflection coefficients of the amplifiers and antenna include multi-path propagation between the measuring instrument and these devices because the measuring instrument will not be a perfect match to the interconnecting cable.  Therefore, we have used a relatively long cable for the measurement so that the frequency structure in the measurement that arises from multi-path propagation would be of relatively finer scale compared to the frequency structure in the reflection coefficient, and hence distinguishable in the measurement.  The measurements have been fitted using a model for the frequency dependence of the reflection coefficients, which does not include---and hence filters out---higher order terms corresponding to multiple reflections between the measuring instrument and device under test.  It is worthwhile stressing that while jointly modeling the observed spectra for sky features and system parameters it is important to use a low-order functional form for the reflection coefficients and not the measured coefficients, because the measured coefficients are contaminated by reflections during the measurements.   The fitted forms are also plotted in Figs.~\ref{fig:refcofM} and \ref{fig:refcofp} along with measurements; deviations in the form of residual ripples exist and are too small to be obvious in the plots shown.

In the SARAS system the balun at the antenna is directly connected to the directional coupler that injects the calibration noise into the signal path; we assume that the dominant path length in the problem is the interconnect between this directional coupler and the cross-over switch, which is integrated with the subsequent components of the receiver.  Reflections over this path length $l$
correspond to a round-trip phase $\Phi = (4 \pi l \nu / f_{\nu} c)$, where $\nu$ is frequency, $c$ is the speed of light in vacuum and 
$f_{\nu}$ is the velocity factor for the propagation of EM signals of frequency $\nu$ in the transmission line.
 
Hereinafter we assume that the gain loss in signal propagation through the paths in the cross-over switch and power splitter are equal:  $g_{\Sigma 1} = g_{\Sigma 2} = g_{\Delta 1} = g_{\Delta 2}=g=(1/\sqrt{2}) $.  

As discussed below, we expect that there will be an unwanted response arising from multi-path propagation within the receiver between the antenna, reference and LNA input ports.  This is predominantly because of mismatch at the antenna terminals; nevertheless, unwanted response also arises from reflections at the reference port.  The reference load is a noise source followed by a precision 23~dB attenuator, which is connected directly to the $J4$ port of the cross-over switch.  The impedance match at this reference port is measured to have a power reflection coefficient less than $-60$~dB.   Therefore, we may expect that the unwanted response arising from reflections at the reference port would be reduced by a factor smaller than $-25$~dB or 0.3\% of the response arising from reflections at the antenna port.  Additionally, because the interconnect to the reference termination is direct and because the precision attenuators and noise source are wide-band devices, the response would be of low order.   Based on the estimates made below on the magnitude of the unwanted response arising from reflections at the antenna port, we estimate that the reflections at the reference port would give rise to a smooth response that is of order 10~mK.  We therefore neglect reflections at the reference port in the analysis presented here.

\subsection{Measurement equations considering reflections to first order} 

In this sub-section, we derive the measurement equations for the cross powers generated by the antenna, reference and the receiver noise voltages.  We consider here a single reflection from the LNA inputs and the antenna terminal. Second and higher order reflections would generate higher harmonics with relatively smaller amplitudes and this is discussed in a later sub-section.

We use symbols $V_a$, $V_{ref}$, $V_{n1}$ and $V_{n2}$ to denote, respectively, voltages representing the antenna noise, reference noise and the noise from the low-noise amplifiers in arms 1 and 2 of the correlation receiver.  $G_{1}$ and $G_{2}$ represent the complex voltage gains of the two receiver chains from the inputs of the low-noise amplifiers onwards. 

\subsubsection{SARAS response to noise power from the antenna}

The calibration noise is injected into the system immediately following the balun via a directional coupler. Therefore, the response of the system is related to antenna noise and calibration noise through the same transfer function. In what follows we compute the response to the antenna temperature $T_{a}$.  When the calibration noise is on, $T_{a}$ may be replaced by $T_{a}+T_{cal}$ to obtain the response.

We write first the signals when the cross-over switch position is `0'.  The direct propagation of antenna signal to the low-noise amplifiers results in voltages $g V_a$ and $-g V_a$ at their input terminals.  The negative sign in the second term explicitly conveys the inversion in signals propagating from $\Delta$ to $P_2$  ports of the power splitter relative to the other paths.  

After reflection from the two receiver inputs, and from the antenna, the additional voltages incident at the two receiver inputs are,
 \begin{equation}
 V_{ar1} = (\Gamma_1+ \Gamma_2) \Gamma_a g^3 V_a e^{i\Phi},
 \end{equation}
and
 \begin{equation}
 V_{ar2} = -(\Gamma_1+ \Gamma_2) \Gamma_a  g^3 V_a e^{i\Phi}.
  \end{equation}
 Adding the direct signals to those after reflection, the net signal voltages at the receiver inputs are
   \begin{equation}
   V_{1} =  g [1+(\Gamma_1+ \Gamma_2) \Gamma_a g^2 e^{i\Phi}]  V_{a} 
   \end{equation}
   and
    \begin{equation}
   V_{2} =  -g [1+(\Gamma_1+ \Gamma_2) \Gamma_a g^2 e^{i\Phi}]  V_{a}.
   \end{equation}

The response of the correlation spectrometer to these voltages is 
\begin{eqnarray}
P_{0a} & = & (G_{1}V_{1}) \times ( G_{2}^*V_{2}^*)  + P_{cor}\nonumber\\
 & = & - g^2 G_{1}G_{2}^* |{[1+(\Gamma_1+ \Gamma_2) \Gamma_a g^2 e^{i\Phi}]}| ^2 V_{a}^2 + P_{cor}.
\end{eqnarray}

At switch position `1', the antenna signal travels to the receiver inputs via the $\Sigma$ port of the power splitter; the reflections also travel via the $\Sigma$ port.  There are no signal inversions and, therefore, the response of the correlation spectrometer to the antenna signal in this case is
\begin{eqnarray}
P_{1a} & = & g^2 G_{1}G_{2}^* |{[1+(\Gamma_1+ \Gamma_2) \Gamma_a g^2 e^{i\Phi}]}| ^2 V_{a}^2 + P_{cor}.
\end{eqnarray}

As discussed earlier in Section~3.2, we difference the measurements recorded in the two switch positions to cancel the spurious correlations ($P_{cor}$) owing to couplings between the arms of the correlation spectrometer; this yields 
\begin{eqnarray}
P_{a}  & = & P_{1a} - P_{0a} \nonumber\\
 & = & 2 g^2 G_{1}G_{2}^*  |{[1+(\Gamma_1+ \Gamma_2)\Gamma_a  g^2 e^{i\Phi}]}| ^2 V_{a}^2\nonumber \\
 & = & 2 g^2 G_{1}G_{2}^* V_{a}^2 [ 1 + |(\Gamma_1+ \Gamma_2)\Gamma_a|^2 g^4 + 
 	2 g^2 {\rm Re}\{(\Gamma_1+ \Gamma_2)\Gamma_a e^{i\Phi}\} ]\nonumber\\
 & = & 2 g^2 G_{1}G_{2}^* V_{a}^2 C_{1}.
\end{eqnarray}
In the last line above we have introduced a new symbol $C_1$ to denote
\begin{equation}
 C_{1} =   1 + |(\Gamma_1+ \Gamma_2)\Gamma_a|^2 g^4 + 
 	2 g^2 {\rm Re}\{(\Gamma_1+ \Gamma_2)\Gamma_a e^{i\Phi}\} .
\end{equation}

When calibrated for the ideal complex gain  $2 g^2 G_{1}G_{2}^*$ of the correlation receiver, the response is wholly real and the imaginary part is expected to be zero.  This response to antenna signal may be viewed as composed of the three terms, all of which are real, in the above expansion for $C_1$.  

The first term (unity) represents the ideal case where no reflections occur.  

\begin{figure}[ht]
\centering
\includegraphics[ angle=270, width=8cm]{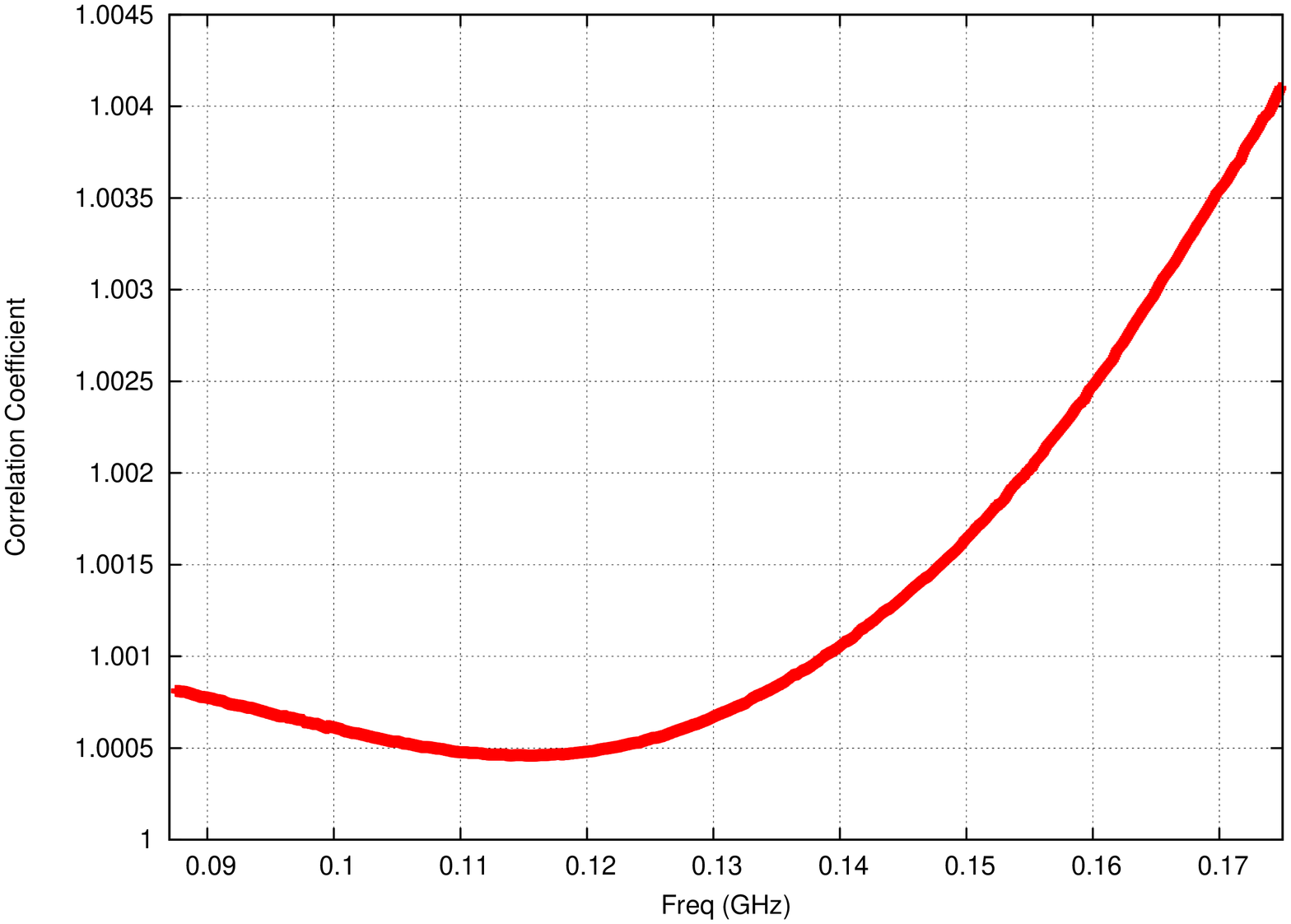}
\caption{SARAS response to antenna temperature: the plot shows the frequency behavior of the 
second term $[1 + |(\Gamma_1+ \Gamma_2)\Gamma_a|^2 g^4]$ in equation (26) that arises owing to products 
between singly reflected voltages along the two receiver chains.}
\label{fig:rr_a}
\centering
\includegraphics[ angle=270, width=8cm]{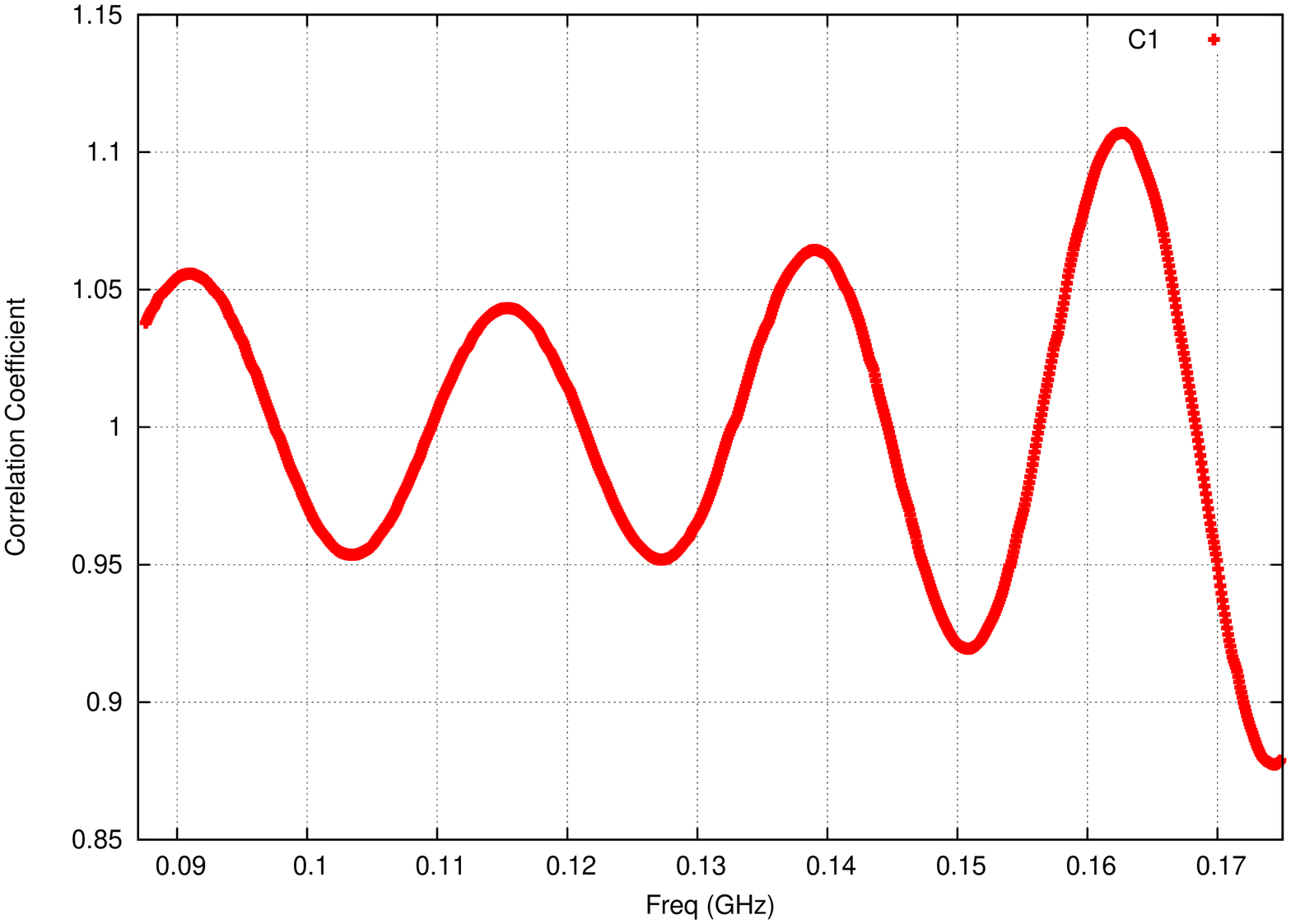}
\caption{Net response to antenna temperature: the plot shows the net response $|{[1+(\Gamma_1+ \Gamma_2)\Gamma_a  g^2 e^{i\Phi}]}| ^2$ that accounts for signals with path delays of at most $2l$.  We have assumed here that the path $l$ corresponding to the cable interconnect is 5~m and the velocity of wave propagation in the cable is $0.6 c$ for the illustration; $c$ is the speed of light in vacuum.}
\label{fig:rd_a}
\end{figure}

The second term is expected to be a slow and smooth function of frequency, essentially dependent on the reflection coefficients at the amplifier inputs and at the antenna.  For the measured coefficients, we show the variation of $[1 + |(\Gamma_1+ \Gamma_2)\Gamma_a|^2 g^4]$ in Fig.~\ref{fig:rr_a}.  This second term represents the additional response arising from the product of signals that arrive at each of the receiver inputs after one reflection and is expected to be 0.05--0.4\% of the antenna temperature or about 0.15--1.2~K.

The third term, which is of larger magnitude than the second, depends on the path length $l$ and would be a sinusoidal ripple across the observing band that is modulated by a relatively slowly varying function of frequency: the modulation amplitude is determined by $2 g^2(\Gamma_1+ \Gamma_2)\Gamma_a$.  This third term arises as a consequence of the product of the direct propagation of the antenna signal to each of the inputs of the low-noise amplifiers with the reflected component that arrives at the other input.  In contrast to the first two terms, this third term is the product of signals that have a relative delay corresponding to twice the length $l$, and hence has a sinusoidal variation of the form sin($4 \pi l \nu / f_{\nu} c$) whose period depends on $l$.  In Fig.~\ref{fig:rd_a} we plot the net response $|{[1+(\Gamma_1+ \Gamma_2)\Gamma_a  g^2 e^{i\Phi}]}| ^2$ assuming that $l=5$~m. The net response is dominated in its frequency structure by the ripple corresponding to the length of the interconnect and is expected to have an amplitude 5--10\% of the antenna temperature, which may be as much as 25~K.  

The analysis illustrates why it is essential to have low reflection coefficients and small transmission line length between the antenna and receiver modules, and devise methods to calibrate for the spectral distortion arising from the inevitable reflections.  
Reflections are systematic departures in the system from ideal behavior.  As we have demonstrated above, these result in the generation of spectral features via coupling with the antenna temperature arising from the dominant foreground.  As we will show in the following sections, unwanted spectral features are also generated via coupling with the reference temperature and receiver noise.   The requirement is that the reflections be reduced to the level where the unwanted and spurious spectral features is smaller than the astronomical signal.  However, this is practically challenging since, for example, the detection of reionization signals requires another 40 dB reduction in power reflection coefficients.  For this reason, we have proposed the hierarchical analysis approach in the paper (see Section~5) wherein the requirement is a minimization of the reflection coefficients to the extent possible followed by characterization and accurate modeling of the observed spectra.  This involves solving for the astronomical parameters by marginalizing over the parameters that describe the reflections.

It may be noted here that even if the two low-noise amplifiers were matched, these unwanted terms would not cancel.  Additionally, since all of the terms multiply $V_{a}^2$, the response to calibration noise power would be simply the same expression except that $V_{a}^2$ is replaced by $V_{cal}^2$. Therefore, calibration by dividing the antenna response by the response to calibration noise would be expected to remove the spurious spectral shapes in the response to antenna temperature, which arise from reflections.

\subsubsection{SARAS response to noise power from the reference}

The response to noise power from the reference may be computed in the same way as above.  In this case the delayed signals arise from reflections of the reference noise from the inputs of the low-noise amplifiers followed by reflections from the antenna.  The transfer function representing the response to the reference would be the same when the reference noise source is on as well as off.

When the switch position is `0', the response of the correlation spectrometer to reference noise voltage $V_{ref}$ is
\begin{equation}
P_{0ref}= g^2 G_{1}G_{2}^*[1+(\Gamma_1- \Gamma_2) \Gamma_a  g^2 e^{i\Phi}] [1-(\Gamma_1- \Gamma_2)  \Gamma_a g^2 e^{i\Phi}]^*  V_{ref}^2 + P_{cor}
\end{equation}
and when the switch position is `1',
\begin{equation}
P_{1ref}= - g^2 G_{1}G_{2}^*[1+(\Gamma_1- \Gamma_2) \Gamma_a   e^{i\Phi}] [1-(\Gamma_1- \Gamma_2) \Gamma_a  g^2 e^{i\Phi}]^* V_{ref}^2 + P_{cor}.
\end{equation}
Differencing the reference power measured in the two switch positions, 
\begin{eqnarray}
P_{ref} & = & P_{1ref} - P_{0ref} \nonumber\\
 & = & - 2 g^2 G_{1}G_{2}^* V_{ref}^2 [ 1 - g^4|(\Gamma_1 - \Gamma_2)\Gamma_a|^2 + i 2 g^2 {\rm Im} \{ (\Gamma_1 - \Gamma_2)\Gamma_a e^{i\Phi} \} ]\nonumber\\
 & = & - 2 g^2 G_{1}G_{2}^* V_{ref}^2 C_{2}.
\end{eqnarray}
As in the previous subsection, we have introduced a new symbol $C_2$ above to denote
\begin{equation}
 C_{2}  =   1 - g^4 |(\Gamma_1 - \Gamma_2)\Gamma_a|^2 + i 2 g^2 {\rm Im} \{ (\Gamma_1 - \Gamma_2)\Gamma_a e^{i\Phi} \}.
\end{equation}
As in the case of the response to antenna temperature, the reference signal components arriving at each of the inputs to the amplifiers consist of directly propagating components---which we refer to hereinafter as $d_1$ and $d_2$---and components that arrive with propagation delays $2l$, which we hereinafter refer to by the symbols $r_1$ and $r_2$.  The first term (which is unity) in the above expression (Equation~24) represents the product $d_1 d_2^*$, the second term represents  $r_1 r_2^*$  and the third term  $d_1 r_2^* +  r_1 d_2^*$.  

\begin{figure}[ht]
\centering
\includegraphics[ angle=270, width=8cm]{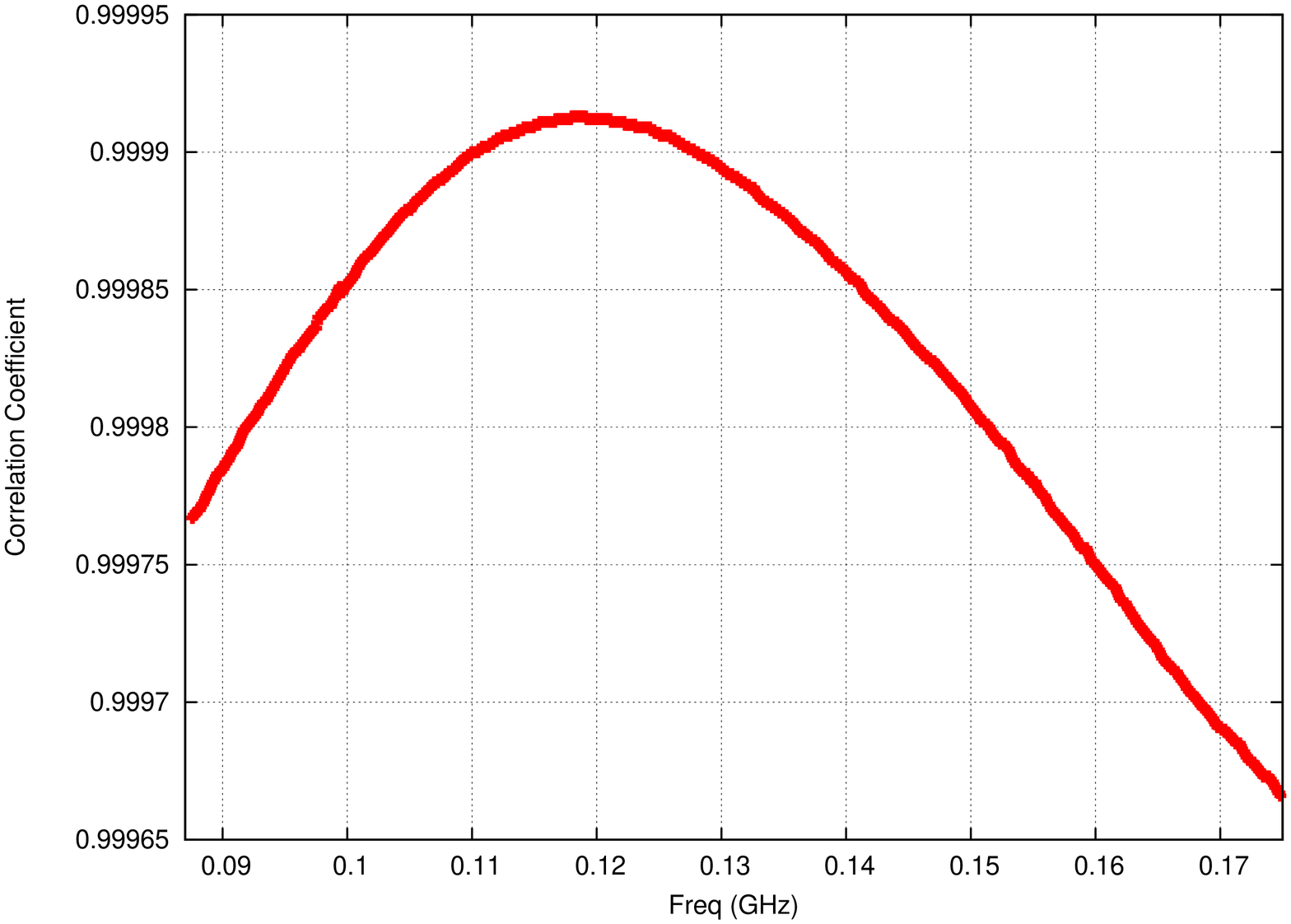}
\caption{Response to noise temperature of the reference: we plot here the term $[1-g^4|(\Gamma_1 - \Gamma_2)\Gamma_a|^2]$ that arises from products between reflected voltages along the two receiver chains.}
\label{fig:rr_r}
\centering
\includegraphics[ angle=270, width=8cm]{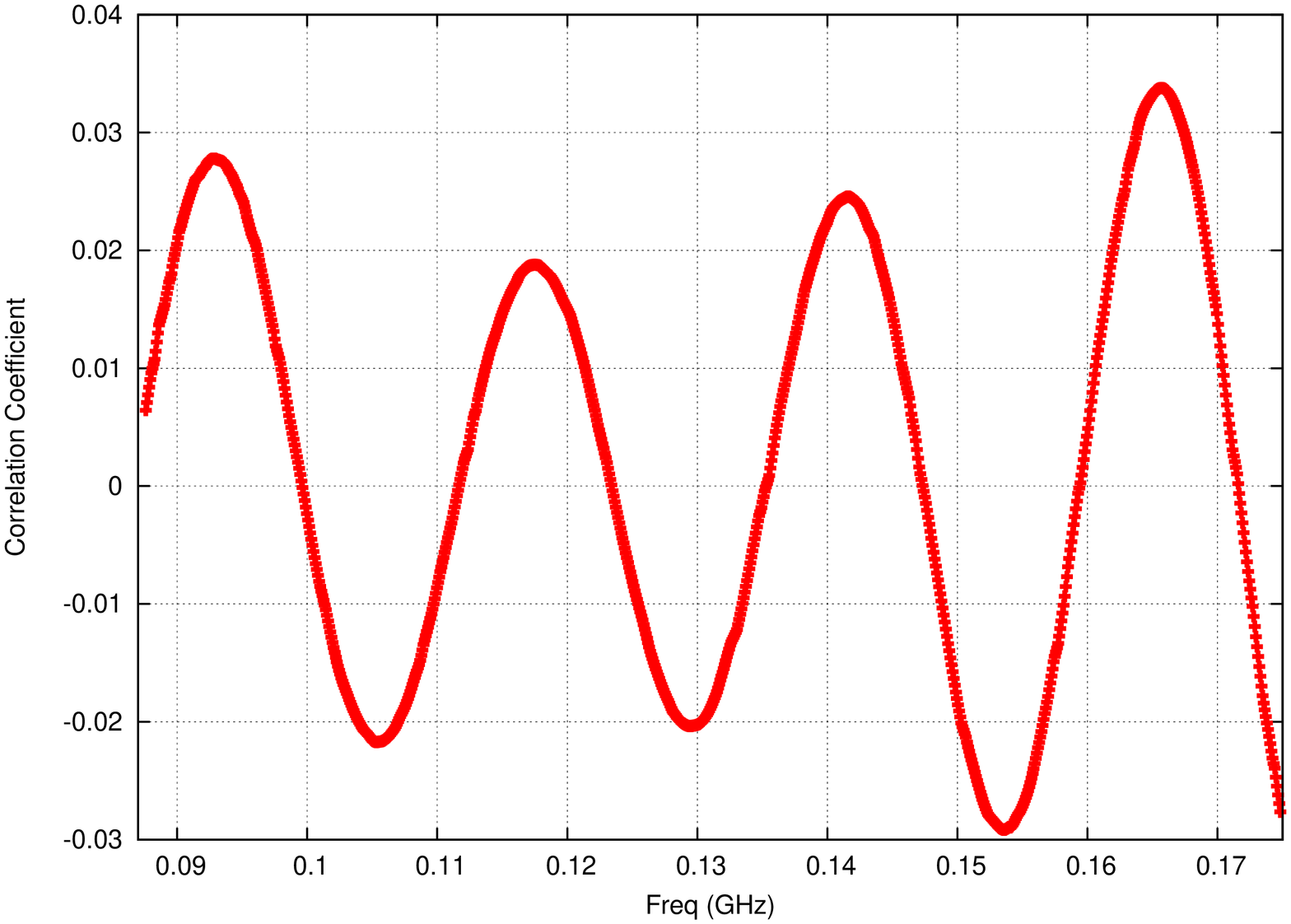}
\caption{Response to noise temperature of the reference: we plot here the term $2 g^2 {\rm Im} \{ (\Gamma_1 - \Gamma_2)\Gamma_a e^{i\Phi} \}$ in the net response that represents the imaginary part.}
\label{fig:rd_r}
\end{figure}

We assume that the response is calibrated for the ideal complex gain $2 g^2 G_{1}G_{2}^*$.  Following such a calibration, the response of the system to reference noise power is complex unlike the response to antenna temperature, which was real. The first two terms in the response in equation~24 constitute the real part where as the third term represents the imaginary part of the complex response.  The first term is what is expected if reflection coefficients were identically zero.  The second term $g^4|(\Gamma_1 - \Gamma_2)\Gamma_a|^2$ is a smooth function of frequency whose spectral structure depends on the reflection coefficients at the amplifiers and antenna.  We plot in Fig.~\ref{fig:rr_r} the first two terms, which constitute the real component of the response, for the measured coefficients in the SARAS system.  The third term, which is imaginary, has a sinusoidal variation with period that depends on the path length $2l$ and has a modulation that depends on $2 g^2 (\Gamma_1 - \Gamma_2)\Gamma_a$.  We plot this imaginary response $2 g^2 {\rm Im} \{ (\Gamma_1 - \Gamma_2)\Gamma_a e^{i\Phi} \}$ in Fig.~\ref{fig:rd_r}, using the measured parameters of the SARAS receiver.

We may make a few remarks here: 

First, in contrast to the case of the response to antenna noise, the unwanted response to reference noise may be eliminated by using a matched pair of low-noise amplifiers.   If $\Gamma_1 = \Gamma_2$, the second and third terms in Equation~24 above vanish.  As discussed above, the corresponding terms in the response to antenna temperature, Equation~20, survive.

As in the case of the response to antenna temperature, the spurious spectral structure arising from the $d_1 r_2^* +  r_1 d_2^*$ term dominates the response.  For the parameters of the SARAS receiver, the $r_1 r_2^*$ response to reference noise is less than 0.1\% of the reference noise temperature where as the ripple in the spectrum owing to the  $d_1 r_2^* +  r_1 d_2^*$ response has an amplitude that is 2--3\% of the reference temperature.  If the reference is a 300~K ambient temperature matched termination, the spurious ripple amplitude is as much as 9~K in amplitude.

The calibration discussed in Section~3.3 envisages a measurement of the response to the calibration step $(P_{cal} - P_{off})$, and a subsequent use of this `bandpass calibration' as a divisor for the response of the system $P_{off}$ so that the quotient would have all ripples associated with reflections of antenna temperature calibrated out.  However, the response $P_{off}$ is a difference measurement that includes the response to the antenna temperature as well as to the reference temperature.   Even if we use matched amplifiers so that the response to the reference would not have any ripple across the bandpass, the calibration process will imprint the ripple, which would be present in the response to the calibration step,  on to the additive response to the reference.  Therefore, as discussed below, the analysis of the measurements of sky spectra must necessarily be a joint  modeling of system parameters along with the antenna power, and is not a simple pipeline calibration of the measured spectra for additive and multiplicative responses to derive the wanted sky spectrum. 

\begin{figure}[ht]
\centering
\includegraphics[ angle=270, width=8cm]{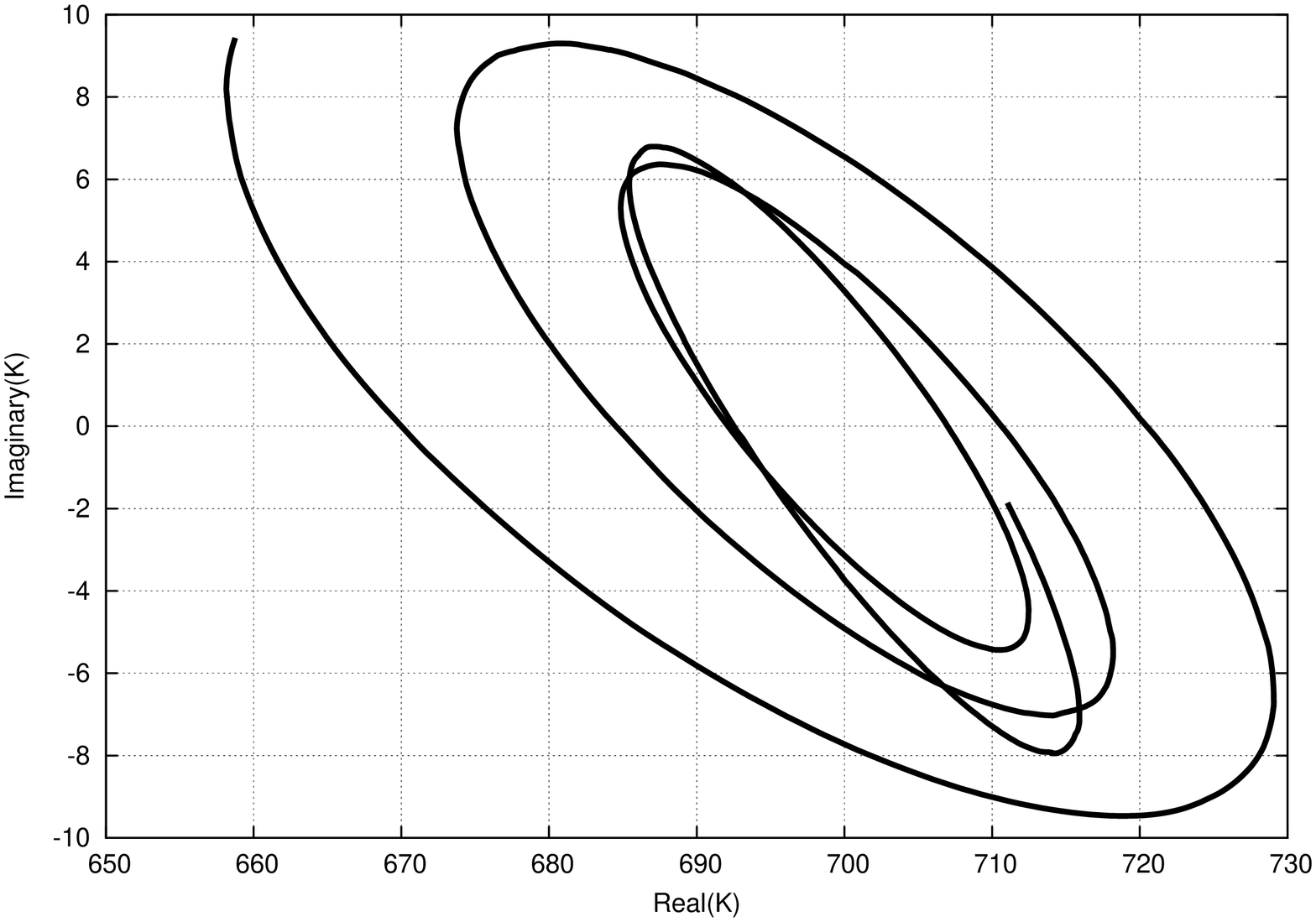}
\centering
\includegraphics[ angle=270, width=8cm]{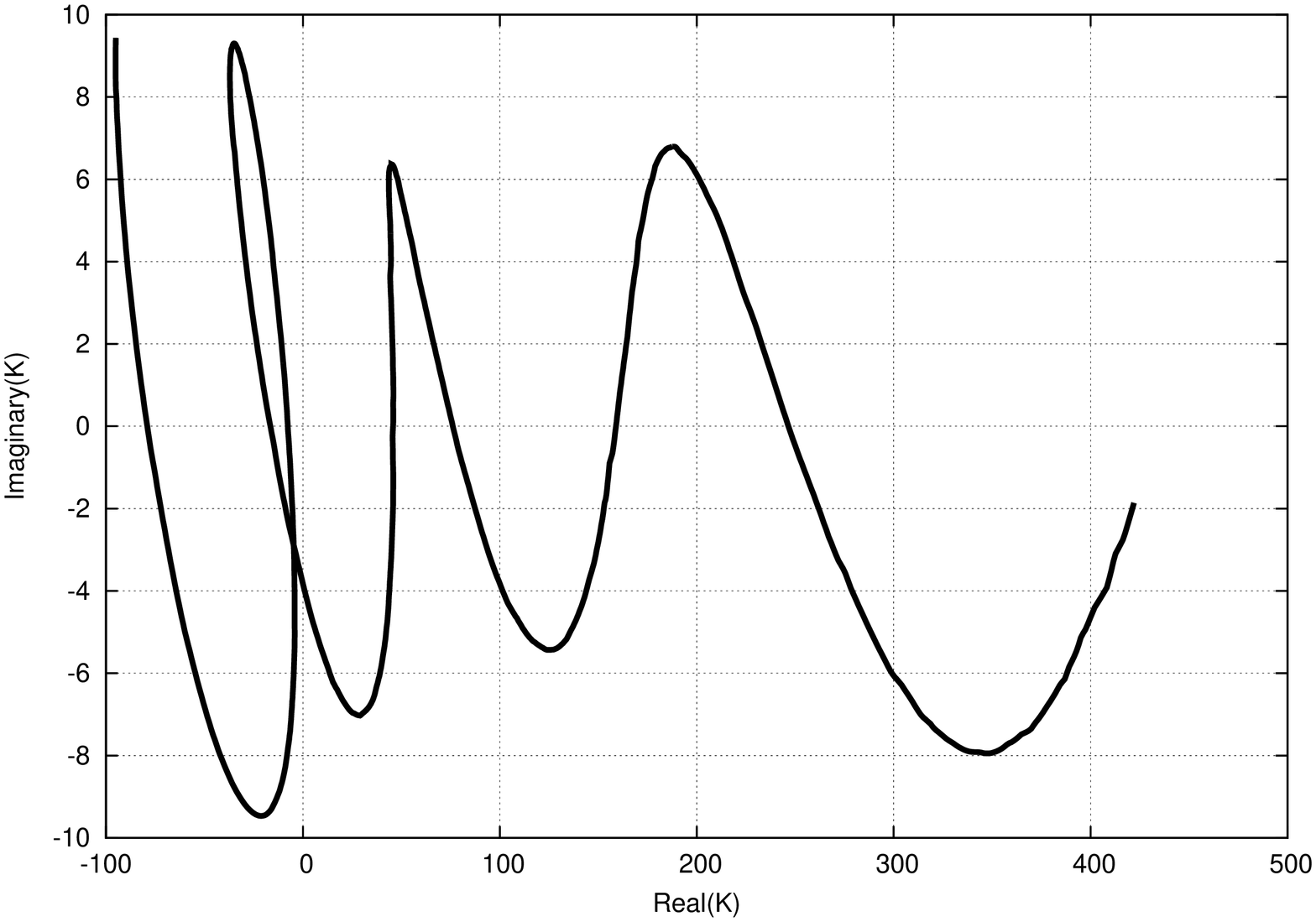}
\caption{Phase diagrams of the differential response to antenna temperature and reference noise.  The reference is assumed to be a 300~K termination.  The figure in the upper panel was obtained by assuming that the antenna temperature is 1000~K; the panel below corresponds to a model antenna temperature that is $[150+200\times(\nu/131.25)^{-2.55}]~K$, where $\nu$ is frequency in MHz. }
\label{fig:arph}
\end{figure}

The differential response to antenna temperature and reference will have structure in the real and imaginary components dominated respectively by modulations with amplitudes proportional to $(\Gamma_1 + \Gamma_2)\Gamma_a$ and $(\Gamma_1 - \Gamma_2)\Gamma_a$; both modulations have the same period in frequency space.  The differential response to antenna and reference temperatures may be displayed as a phase diagram in the complex plane as shown in Fig.~\ref{fig:arph}.  

 \subsubsection{SARAS response to receiver noise}

 Amplifier noise has a component that adds to the input signal and propagates downstream along with the amplified input signal; it has a second component---correlated with the first---that propagates upstream against the nominal signal flow direction. A part of these upward traveling receiver noise components from both low-noise amplifiers are reflected at the antenna (as well as at the reference if an impedance mismatch is encountered at that port) and would propagate downstream to both amplifiers and add, with a delay, to the receiver noise that originally propagated downstream. This causes mutual coupling of receiver noise between the two arms of the correlation spectrometer.
 
Let $f_{1}$ and $ f_{2}$ be the fractions of the noise voltages of amplifiers 1 and 2 that propagate in reverse direction.  We denote the amplifier noise voltages by $V_{n1}$ and $V_{n2}$; the reverse propagating noise voltages are $f_1 V_{n1}$ and $f_2 V_{n2}$.  Herein we express the reverse propagating noise powers as fractions of the usual receiver noise power and the reference point is the inputs of the low-noise amplifiers.
 
In this section, we consider the usual noise powers of the amplifiers plus additional components arising from a single dominant reflection of backward propagating receiver noise, the reflection is assumed to happen at the antenna.  At switch position `0', the net amplifier noise voltages at the two amplifier inputs are:
  \begin{equation}
  V_{nr1} =  V_{n1} + (f_{1}V_{n1} - f_{2}V_{n2}) \Gamma_a g^2 e^{i\Phi},
   \end{equation}
   and
   \begin{equation}
  V_{nr2} = V_{n2} + (f_{2}V_{n2} - f_{1}V_{n1}) \Gamma_a g^2 e^{i\Phi}.
   \end{equation}
The noise voltages $V_{n1}$ and $V_{n2}$ are uncorrelated and the correlation spectrometer gives zero mean response to terms that involve products between these voltages.  Omitting such terms, the response of the correlation spectrometer to amplifier noise may be written as
\begin{eqnarray} \nonumber
P_{0n}  & = & (G_{1}V_{nr1}) \times (G_{2}^* V_{nr2}^*)  + P_{cor}\nonumber\\
 & = & - G_{1}G_{2}^{*} [ g^4 \Gamma_a^2 \{ f_1^2 V_{n1}^2 + f_2^2 V_{n2}^2 \} \nonumber \\
 &&    + g^2 \{ V_{n1}^2 f_1^* \Gamma_a^* e^{-i\Phi} + V_{n2}^2 f_2 \Gamma_a e^{i\Phi} \} ] + P_{cor}.
\end{eqnarray}   
Similarly, we may write the response in switch position `1' as
\begin{eqnarray}\nonumber
P_{1n} & = & G_{1}G_{2}^{*} [ g^4 \Gamma_a^2 \{ f_1^2 V_{n1}^2 + f_2^2 V_{n2}^2 \}\nonumber  \\
    && + g^2 \{ V_{n1}^2 f_1^* \Gamma_a^* e^{-i\Phi} + V_{n2}^2 f_2 \Gamma_a e^{i\Phi} \} ] + P_{cor}.
\end{eqnarray}
Differencing the responses to receiver noise power in the two switch positions, 
\begin{eqnarray}
P_{n} & = & P_{1n} - P_{0n} \nonumber \\
 & = & 2 g^2 G_{1}G_{2}^{*} [ g^2 \Gamma_a^2 \{ f_1^2 V_{n1}^2 + f_2^2 V_{n2}^2 \} 
  + \{ V_{n1}^2 f_1^* \Gamma_a^* e^{-i\Phi} + V_{n2}^2 f_2 \Gamma_a e^{i\Phi} \} ]\nonumber\\
 & = & 2 g^2 G_{1}G_{2}^{*} [ g^2 \Gamma_a^2 f_1^2  + f_1^* \Gamma_a^* e^{-i\Phi} ] V_{n1}^2
  +2 g^2 G_{1}G_{2}^{*} [ g^2 \Gamma_a^2 f_2^2  + f_2^* \Gamma_a^* e^{-i\Phi} ] V_{n2}^2\nonumber\\
& = & 2 g^2 G_{1}G_{2}^{*} [C_{n1} V_{n1}^2
  +C_{n2} V_{n2}^2].
\end{eqnarray}
Once again we introduce new symbols $C_{n1}$ and $C_{n2}$ that are
\begin{equation}
 C_{n1} =     [ (g f_1 \Gamma_a )^2  +  (f_1 \Gamma_a e^{i\Phi})^{*}  ]
\end{equation}
and
\begin{equation}
 C_{n2} =     [ (g  f_2 \Gamma_a )^2  +  (f_2 \Gamma_a e^{i\Phi})  ].
\end{equation}

Following the notation adopted in previous subsections, the first terms in Equations~30 \& 31 are $r_1 r_2^*$ type terms and the later are $d_1 r_2^*$ and $r_1 d_2^*$ type terms.  The first terms are real and subdominant; they are smoothly varying functions of frequency that depend on the frequency dependance of the receiver noise, that of fractions $f_{1}$ and $ f_{2}$, and that of $\Gamma_a$.  

Assuming that the fractions $f_{1}$ and $ f_{2}$ are real quantities, and writing $\Gamma_a = | \Gamma_a | e^{-\Phi_a}$, the second terms in Equations~30 \& 31 may be expanded together as follows:
\begin{eqnarray}
V_{n1}^2 f_1 \Gamma_a^* e^{-i\Phi} + V_{n2}^2 f_2 \Gamma_a e^{i\Phi} \nonumber \\
 & = & V_{n1}^2 f_1 |\Gamma_a| e^{-i(\Phi + \Phi_a)} + V_{n2}^2 f_2 |\Gamma_a| e^{i(\Phi + \Phi_a)} \nonumber \\
 & = & |\Gamma_a| [ ( f_1 V_{n1}^2 + f_2 V_{n2}^2) {\rm cos} (\Phi + \Phi_a) - \nonumber \\
 & & i ( f_1 V_{n1}^2 - f_2 V_{n2}^2) {\rm sin} (\Phi + \Phi_a) ].
\end{eqnarray}
This is the dominant term representing the summation of $d_1 r_2^*$ and $r_1 d_2^*$ type terms and this part of the response is complex.  The real and imaginary parts are both modulated sinusoids with $\pi/2$ phase difference, whose amplitudes are proportional to $( f_1 V_{n1}^2 + f_2 V_{n2}^2)$ and $( f_1 V_{n1}^2 - f_2 V_{n2}^2)$.   In the complex plane, the trace of the response is ellipses with major and minor axes $|\Gamma_a|( f_1 V_{n1}^2 + f_2 V_{n2}^2)$ and $|\Gamma_a|( f_1 V_{n1}^2 - f_2 V_{n2}^2)$ respectively.

We have experimentally measured the fraction of the receiver noise that travels upwards in the SARAS system.  To reconfigure for this measurement, we first remove the antenna as well as the switch and power-splitter assembly.  While one LNA  input is terminated with a matched impedance,  the LNA under test has a cable of length $l = 5$~m connecting its input  directly to the output of the directional coupler.  The main line input of the directional coupler is terminated in an `open' load and the coupled port is terminated with a matched impedance. The auto-correlation spectrum of the output of the LNA under test is recorded. In this configuration the fractional noise from the LNA, which travels upstream, suffers total internal reflection from the open load. Therefore, the measured auto-correlation has contributions from the receiver noise that travels directly downstream, and a contribution arising from the additional correlations owing to the reflected receiver noise. Since the direct and the reflected receiver noise voltages have a phase difference corresponding to the path difference of $2l$, the correlation spectrum is oscillatory in nature. We write below the analysis for the LNA with label `1'.  Following the same notation used above the net noise voltage at the input of the LNA, considering one reflection of the upward traveling receiver noise at the open load, is given by:
\begin{equation}
V_{nr1} = V_{n1}+f_{1}V_{n1}e^{i\Phi}.
\end{equation} 
The power spectrum corresponding to the noise voltage at the LNA may be written as:
\begin{eqnarray}
P_{n1}  & = & (G_{1}V_{nr1}) \times (G_{1}^{*}V_{nr1}^{*}) \nonumber \\
& = & G_{1}^2 |V_{nr1}|^2 \nonumber \\
& = & G_{1}^2 V_{n1}^2(1 + f_{1} ^2  +2  {\rm Re} \{f_{1} e^{i\Phi}\}).
\end{eqnarray}
The first term in parenthesis (unity) represents the autocorrelation of the receiver noise traveling directly downstream: $d_1d_1^*$. The second term is  the autocorrelation of the upward traveling receiver noise that is reflected back from the open load: $r_1r_1^*$. The third term represents the correlation $(d_1r_1^* + d_1^*r_1)$ between the downward traveling receiver noise and its upward traveling component, following reflection.  

Bandpass calibration of the above measurement is done in a modified setup.  The 5-m cable is omitted, the `open' load is replaced with a matched termination, and a calibration noise source is connected to the coupled port.  The calibration noise is toggled between `on' and `off' every 0.7~s and the bandpass calibration is derived by subtracting the recorded power spectra in `off' state from that in `on' state. The bandpass calibration is given by:
\begin{equation}
P_{cal} = G_{1}^2 V_{cal}^2. 
\end{equation}   
Dividing $P_{n1}$ by $P_{cal}$ and multiplying by the temperature  $T_{n1}$ of the calibration noise source, the calibrated spectrum of the power from receiver 1 is given by:
\begin{equation}
T_{11} = T_{n1} (1 + f_{1}^2 +2{\rm Re} (f_{1} e^{i\Phi})).
\end{equation} 

Higher order reflections of LNA amplifier noise between the LNA input and the open end of the transmission line would add signals with delays that are integral multiples of $2 l$.  These would generate the ripples in the measured spectrum with higher harmonics, amplitudes of which are successively reduced by about 10\%.  We estimate the fractions $f_1$ and $f_2$ via fits to the measured spectra for only the fundamental; therefore, we have neglected writing the analysis above as a series expansion representing the higher order reflections.

We fit the measured data to the form $a_{0}+a_{1}{\rm sin}(a_{2}f)$, where $a_{0}$ represents the mean receiver noise power and $a_{1}$ represents the amplitude of the ripple. The residual to the fit contains higher order reflections as well as a residual ripple due to the frequency dependence of cable loss and reflection coefficient at the LNA input.  Following calibration, the measured spectra corresponding to the two receivers are shown below in Fig.~\ref{fig:rec_noise} along with the fits.  We thus obtain $f_1=9.9\%$ and $f_2 = 4.67\%$.  

\begin{figure}[ht]
\centering
\includegraphics[ angle=270, width=8cm]{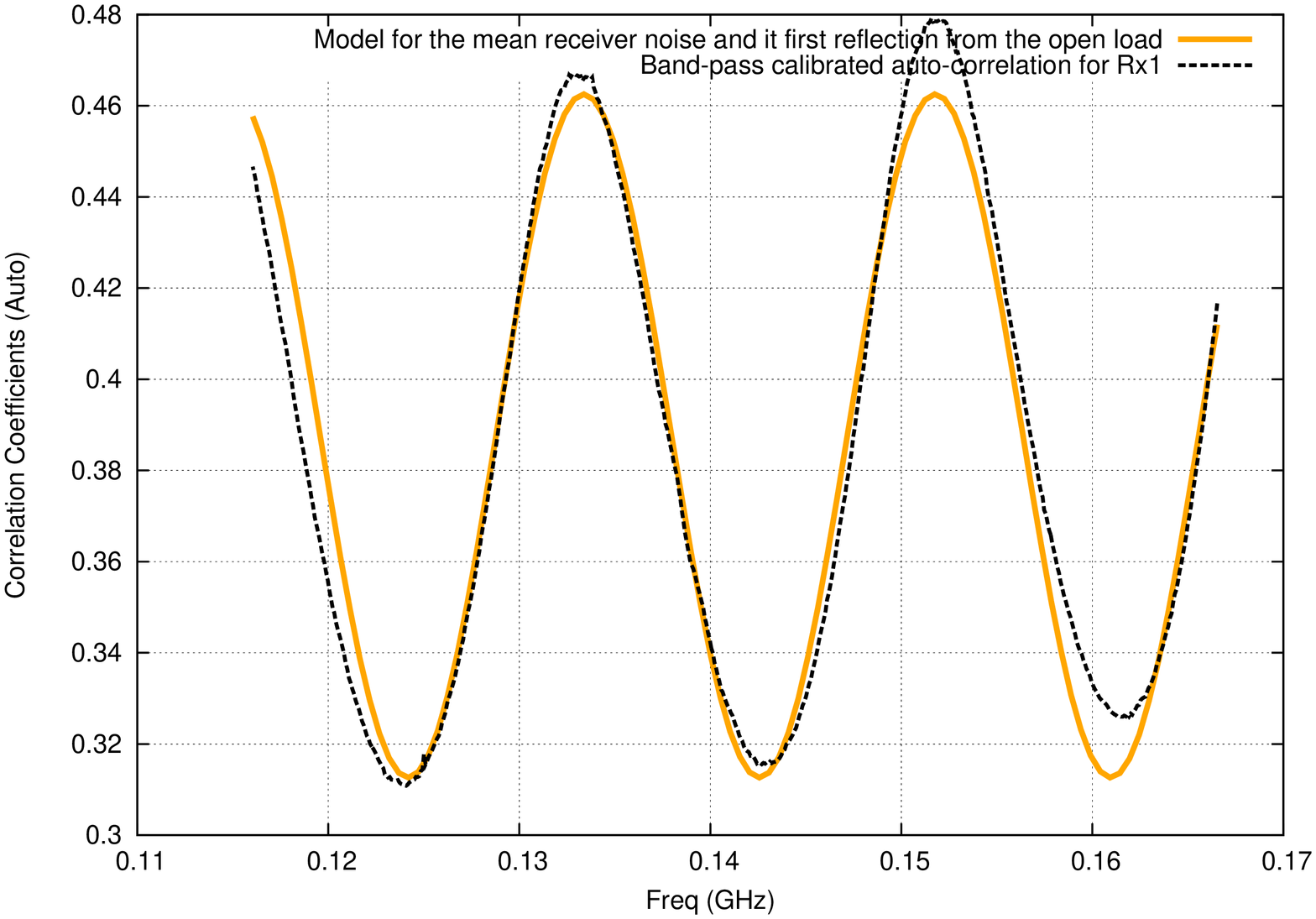}
\centering
\includegraphics[ angle=270, width=8cm]{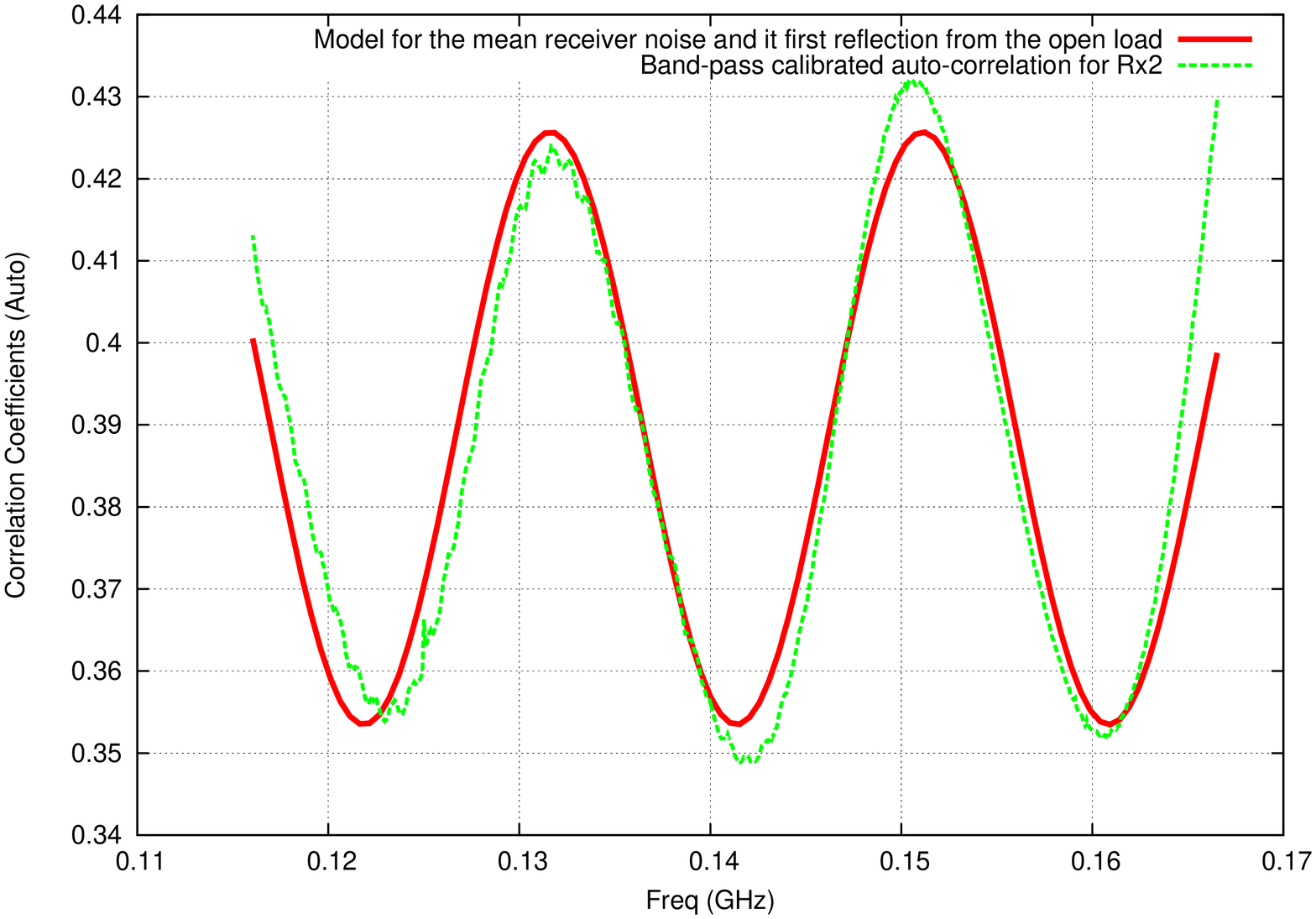}
\caption{Calibrated spectra of the power from the receivers with 5~m cables connected to their inputs; the cables are open at their ends to reflect the amplifier noise traveling out of the input ports.  Also shown are fits using sinusoidal forms. The upper panel is for the first receiver chain and the lower panel is for the second chain. The y-axis is in arbitrary units.}
\label{fig:rec_noise}
\end{figure}

\begin{figure}
\centering
\includegraphics[ width=6cm, angle=-90]{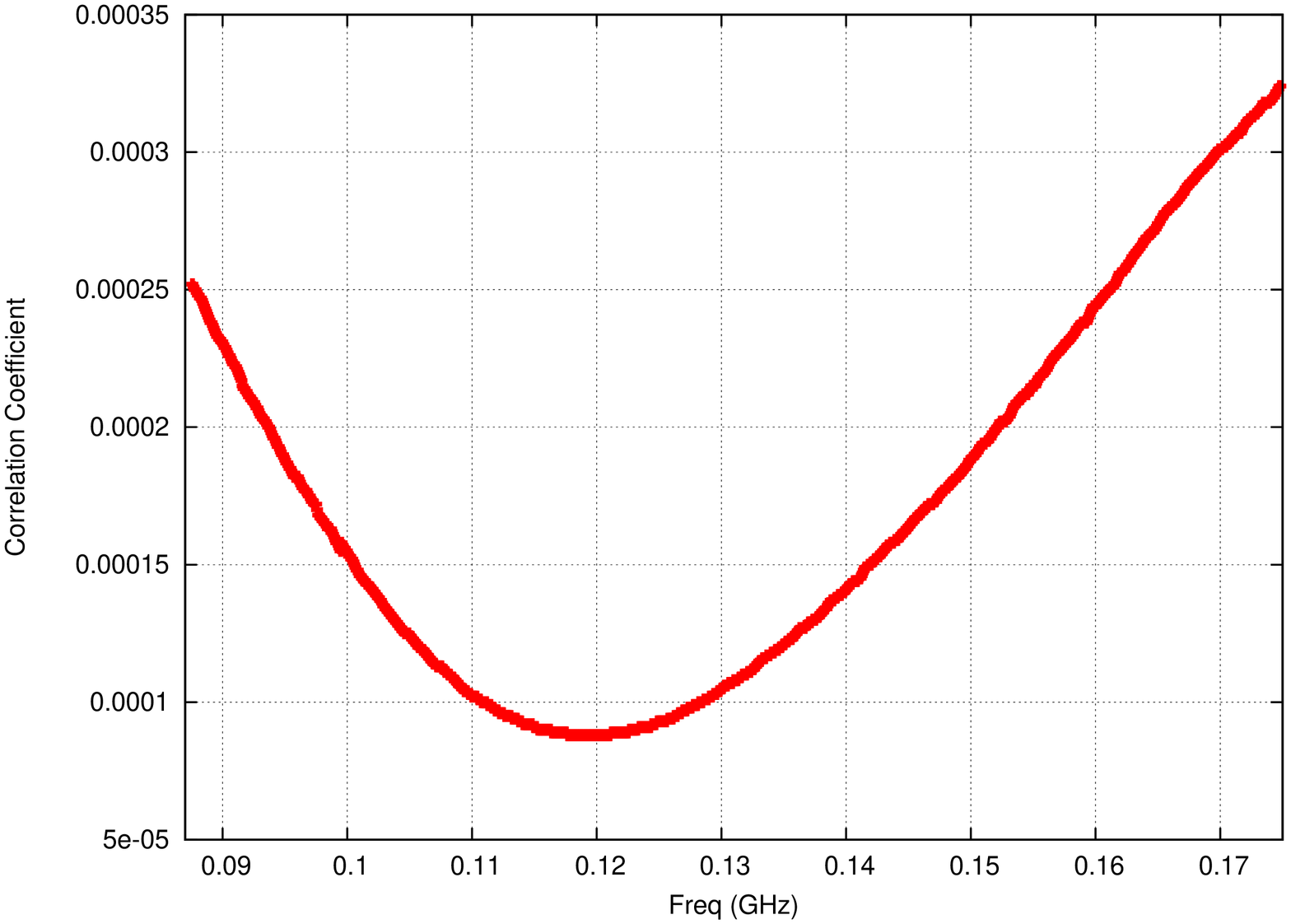}
\caption{Response to receiver noise: the figure shows the term $g^2 \Gamma_a^2 \{ f_1^2  + f_2^2 \} $ in Equation~35 that arises from $r_1 r_2^*$ products between reflected signals of upward traveling receiver noise voltages from the two low-noise amplifiers.}
\label{fig:rr_rn}       
\end{figure}

\begin{figure}
\centering
\includegraphics[ angle=270, width=7.2cm]{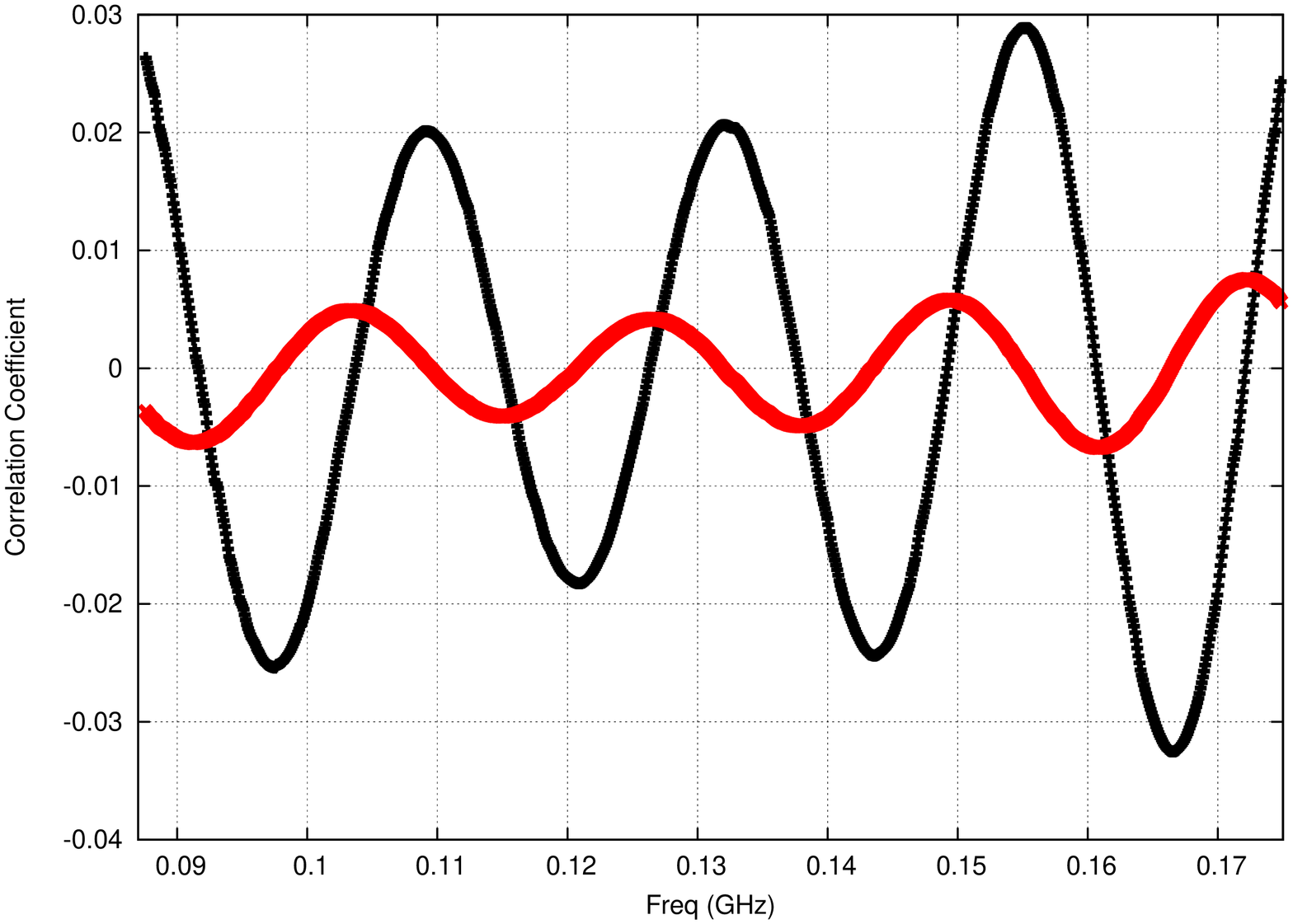}
\centering
\includegraphics[ angle=270, width=7.2cm]{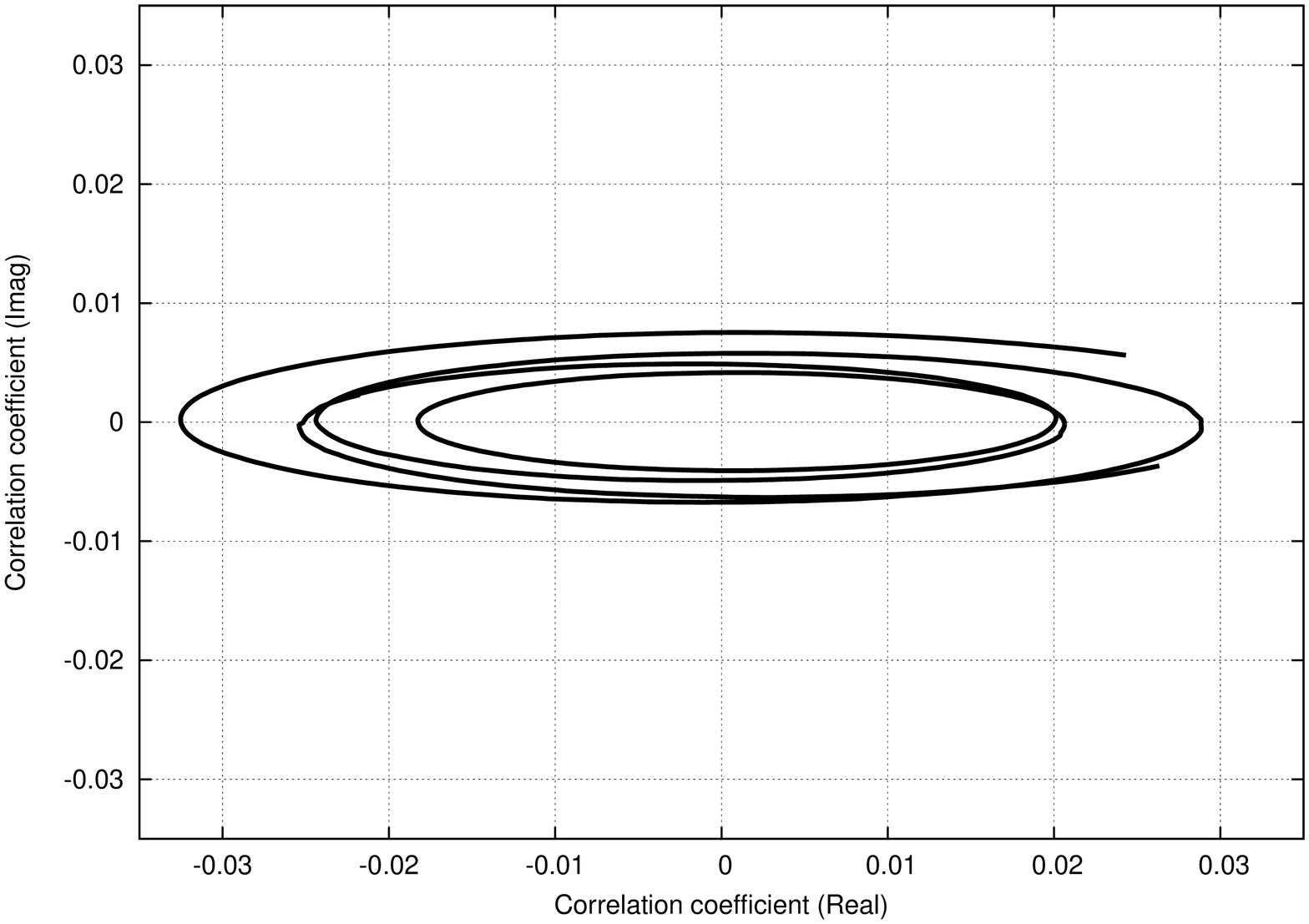}
\caption{The complex response to receiver noise; what is plotted is the sum of factors that multiply the receiver noise to yield the response.  On the top panel is shown the real and imaginary components of the response; in the lower panel the response is shown in the complex plane.}
\label{fig:rn}
\end{figure}

We plot  $g^2 \Gamma_a^2 \{ f_1^2  + f_2^2 \} $ for the SARAS system in Fig.~\ref{fig:rr_rn} adopting the above measured values for the fractions $f_{1}$ and $ f_{2}$.  The sum of these terms in Equations~30 \& 31 is expected to manifest in a response with magnitude less than 10~mK for receiver noise temperatures of a few tens of Kelvin.  For $l=5$~m, we plot in Fig.~\ref{fig:rn} the real and imaginary components $|\Gamma_a| ( f_1 + f_2 ) {\rm cos} (\Phi + \Phi_a) $ and $|\Gamma_a| ( f_1 - f_2 ) {\rm sin} (\Phi + \Phi_a)  $ in the upper panel.  The complex response is also shown as a phase diagram in the complex plane in the lower panel of the figure.   

The response may have an amplitude as much as 5\% of the receiver noise temperature.  The imaginary response is always less than the real and is offset $\pi/2$ in phase.  If the amplifiers form a matched pair and $f_1 = f_2$ and $ \langle V_{n1}^2 \rangle = \langle V_{n2}^2 \rangle $, the imaginary term vanishes.

\subsection{Measurement equations considering higher order reflections within the system}

The analysis described above yields the response of the SARAS spectrometer considering only first order reflections.  We have therein included effects arising from a single reflection at the antenna. However, precision measurements of the cosmic radio background with realistic systems, which aim to detect weak features from the epoch of reionization, would require a more complete analysis that takes into account multiple or higher order reflections of the system temperature components. The higher order reflections generate ripples of higher order in the spectra measured by the correlation spectrometer;  these would be at harmonics of the fundamental ripple discussed in previous sections.  Herein we develop the measurement equations that include effects of higher order reflections in the form of series expansions.

\subsubsection{SARAS response to noise power from the antenna}

We begin by considering the response to the signal from the antenna.  Considering the state of the system with cross-over switch in position `0', the net signal voltages at the pair of amplifier inputs are
 \begin{eqnarray} \nonumber
   V_{1} & =  & g [1+(\Gamma_1+ \Gamma_2) \Gamma_a g^2 e^{i\Phi} 
   + (\Gamma_1+ \Gamma_2)^2 \Gamma_a^2 g^4 e^{i2\Phi} +  ...........  \\
   && ........ + (\Gamma_1+ \Gamma_2)^n \Gamma_a^n g^{2*n} e^{in\Phi} + ..........]  V_{a} 
 \end{eqnarray}
   and
\begin{eqnarray} \nonumber
   V_{2} & = & -  g [1+(\Gamma_1+ \Gamma_2) \Gamma_a g^2 e^{i\Phi} 
   	+ (\Gamma_1+ \Gamma_2)^2 \Gamma_a^2 g^4 e^{i2\Phi} +  ........... \\ 
	&& ........... + (\Gamma_1+ \Gamma_2)^n \Gamma_a^n g^{2*n} e^{in\Phi} + ..........]  V_{a}.
   \end{eqnarray}
Using the symbol $\Upsilon$ to denote the expression $(\Gamma_1+ \Gamma_2) \Gamma_a g^2$,
\begin{eqnarray}\nonumber
  V_{1} & = & g [1+ \Upsilon e^{i\Phi} + \Upsilon^{2} e^{i2\Phi} +  ....+ \Upsilon^n e^{in\Phi} + ..........]  V_{a} \\
   & = &  g(\displaystyle\sum\limits_{m=0}^{\infty}  \Upsilon^{m}  e^{i m\Phi}) V_{a}
\end{eqnarray}
and, similarly, 
 \begin{equation}
   V_{2} =- g(\displaystyle\sum\limits_{m=0}^{\infty}  \Upsilon^{m}  e^{i m\Phi}) V_{a}.
  \end{equation}
The net response of the correlation spectrometer for this switch position may be written as
 \begin{eqnarray} \nonumber 
P_{0a} & = & (G_{1}V_{1}) \times (G_{2}V_{2})^* +P_{cor}\\
             & = & -g^{2}G_{1}G_{2}^* (\displaystyle\sum\limits_{p=0}^{\infty}  \Upsilon^{p}  e^{ip\Phi})  (\displaystyle\sum\limits_{q=0}^{\infty}  \Upsilon^{q}  e^{iq\Phi})^* V_{a}^2 +P_{cor} \nonumber \\
           &  = &  -g^{2}G_{1}G_{2}^* [ \displaystyle\sum\limits_{m=0}^{\infty}  |\Upsilon^{m}|^{2} (1+ {\rm Re} (\Upsilon e^{i\Phi}) + {\rm Re} (\Upsilon^2 e^{i2\Phi}) \nonumber \\
           && +....+{\rm Re} (\Upsilon^n e^{in\Phi}) + ..........) ] V_{a}^2 +P_{cor}.
\end{eqnarray}
At switch position `1', $(P_{1a} - P_{cor}) = -(P_{0a} - P_{cor})$. Subtracting the cross-power in the two switch positions to cancel $P_{cor}$, we get
\begin{eqnarray}
P_{a} & = &  P_{1a} - P_{0a} \nonumber \\
 & = & 2g^{2}G_{1}G_{2}^* [\displaystyle\sum\limits_{m=0}^{\infty}(|\Upsilon^{m}|^{2}) 
\displaystyle\sum\limits_{k=0}^{\infty} {\rm Re} (\Upsilon^k e^{ik\Phi})] V_{a}^2. \nonumber \\
&=&  2g^{2}G_{1}G_{2}^* [\displaystyle\sum\limits_{m=0}^{\infty}(|\Upsilon^{m}|^{2}) ] V_{a}^2 + \nonumber \\
&& 2g^{2}G_{1}G_{2}^* [ {\rm Re} (\Upsilon e^{i\Phi}) \displaystyle\sum\limits_{m=0}^{\infty}(|\Upsilon^{m}|^{2}) ] V_{a}^2 + \nonumber \\
&& 2g^{2}G_{1}G_{2}^* [ {\rm Re} (\Upsilon^2 e^{i2\Phi}) \displaystyle\sum\limits_{m=0}^{\infty}(|\Upsilon^{m}|^{2}) ] V_{a}^2 + \nonumber \\
&& ........... + 2g^{2}G_{1}G_{2}^* [ {\rm Re} (\Upsilon^n e^{in\Phi}) \displaystyle\sum\limits_{m=0}^{\infty}(|\Upsilon^{m}|^{2}) ] V_{a}^2 + ...........
\end{eqnarray}

The terms in the above expansion may be interpreted as follows. The cross correlation of signals arriving directly at the two LNA inputs, as well as of signals arriving at the pair of inputs after having undergone identical number of reflections, would result in spectral responses that would vary smoothly as a function of frequency: these are the terms in the first line of the expansion. However, every additional path delay of  $2l$ in which the signals reflect at an LNA input and antenna cause the incident voltage amplitude at the LNA input to reduce by factor $\Upsilon = (\Gamma_1+ \Gamma_2) \Gamma_a g^2$, which is at most  5\%. The cross correlation response to components that have undergone successive reflections reduces by factor $\Upsilon^2 =  ((\Gamma_1+ \Gamma_2) \Gamma_a g^2)^2$, which is about 0.3\% of the response to the adjacent lower order reflection.  Beyond the first order reflection, high order terms yield responses below about 1~mK; therefore, we may neglect their contributions to the net response of the system. 

The second line in the expansion represents products between antenna signals that are reflected $p$ and $p+1$ times.  The correlation between any two successive reflected components would result in a spectral ripple with the fundamental period. However, in the response the amplitudes of such ripples in successive terms would again be reduced by factors $\Upsilon^2$.  The first term in this series may be as large as 25~K,  the second term would be at most 75~mK in amplitude and later terms less than 1~mK.  

The third line in the expansion represents the first harmonic, which arises from correlations between antenna signals arriving at the LNA inputs with relative delays of $4l$.  The first term with this first harmonic ripple period merits consideration since its amplitude is only a factor $|\Upsilon|$ (which could be as much as 5\%) below the amplitude of the fundamental.  The second term is an additional factor $\Upsilon^2$ lower and hence would not be more than a few mK.

The correlation between direct component along one path with the reflected component along the other path which has undergone `$n$' reflections would result in a ripple which is $(n-1)$'th harmonic of the fundamental ripple frequency.  This $(n-1)$'th harmonic has an amplitude that is a factor $|\Upsilon|^{n-1}$ of the fundamental.  In the SARAS system, the third and higher harmonics would have amplitudes below a few mK and may be neglected while modeling EoR measurements.  Thus, for the modeling of SARAS spectral measurements that are aimed at detecting the all-sky EoR signatures, the response to antenna noise power may be approximated as:
\begin{equation}
P_{a} \approx   2g^{2}G_{1}G_{2}^* [  1 +  |\Upsilon|^{2} +  {\rm Re} (\Upsilon e^{i\Phi}) ( 1 +  |\Upsilon|^{2})   +   {\rm Re} (\Upsilon^2 e^{i2\Phi}) ( 1 +  |\Upsilon|^{2})  ] V_{a}^2.
\end{equation}

\subsubsection{SARAS response to noise power from the reference}

In this subsection we analyze the response of the SARAS system to noise temperature of the reference.  As above, we consider only reflections from the antenna terminal and neglect reflections from the reference terminals where the impedance of the termination is adequately matched.  
 In case of the antenna temperature, successive reflections of the signal from the antenna to the LNAs traverse the same path as that of the direct incidence.   However, in the case of the reference noise voltage the multiply reflected signals travel paths between the antenna and LNA inputs and not the path of direct incidence between the reference and LNA inputs.  This asymmetry results in that the response to reference noise power is dominated by a Fourier series of ripples in the imaginary component of the complex response, unlike the case for the response to antenna noise power that is dominated by a Fourier series of ripples in the real part.  
 
 We begin by considering the response of the system with cross-over switch in position `0'.   The notations used below are the same as in Section~4.2.1.  The net signal voltages at the pair of amplifier inputs are
 \begin{eqnarray}
   V_1 & =  & g [1+(\Gamma_1- \Gamma_2) \Gamma_a g^2\{e^{i\Phi} +(\Gamma_1+ \Gamma_2) \Gamma_a g^2 e^{2i\Phi} 
   + (\Gamma_1+ \Gamma_2)^2 \Gamma_a^2 g^4 e^{i3\Phi} +  \nonumber \\
   && ........ + (\Gamma_1+ \Gamma_2)^{(n-1)} \Gamma_a^{(n-1)} g^{2(n-1)} e^{in\Phi} + ...........\}]  V_{ref} 
 \end{eqnarray}
   and
 \begin{eqnarray} 
   V_2& =  & g [1-(\Gamma_1- \Gamma_2) \Gamma_a g^2\{e^{i\Phi} +(\Gamma_1+ \Gamma_2) \Gamma_a g^2 e^{i2\Phi} 
   + (\Gamma_1+ \Gamma_2)^2 \Gamma_a^2 g^4 e^{i3\Phi} +  \nonumber  \\
   && ........ + (\Gamma_1+ \Gamma_2)^{(n-1)} \Gamma_a^{(n-1)} g^{2(n-1)} e^{in\Phi} + ........... \}]  V_{ref}.
 \end{eqnarray}
 We continue to use the symbol $\Upsilon$ to denote $(\Gamma_1+ \Gamma_2) \Gamma_a g^2$, introduce an additional symbol $\Psi$ to denote $(\Gamma_1- \Gamma_2) \Gamma_a g^2$, and reduce the above equations to
\begin{eqnarray}\nonumber
  V_{1} & = & g [1+ \Psi \{e^{i\Phi}+ \Upsilon e^{i2\Phi} + \Upsilon^{2} e^{i3\Phi} +  ....+ \Upsilon^{(n-1)} e^{in\Phi} + ........... \}]  V_{ref} \\
   & = &  g[1+ \Psi \{\displaystyle\sum\limits_{m=1}^{\infty}  \Upsilon^{(m-1)}  e^{i m\Phi}\}] V_{ref}
\end{eqnarray}
and, similarly, 
 \begin{eqnarray}
  V_{2} & = &  g[1- \Psi \{\displaystyle\sum\limits_{m=1}^{\infty}  \Upsilon^{(m-1)}  e^{i m\Phi}\}] V_{ref}.
\end{eqnarray}
The net response of the correlation spectrometer for this switch position may be written as
 \begin{eqnarray} 
P_{0ref} & = & (G_{1}V_{1}) \times (G_{2}V_{2})^* +P_{cor} \nonumber  \\
                & = & g^{2}G_{1}G_{2}^* [1+ \Psi (\displaystyle\sum\limits_{p=1}^{\infty}  \Upsilon^{(p-1)}  e^{i p\Phi})] 
   [1- \Psi (\displaystyle\sum\limits_{q=1}^{\infty}  \Upsilon^{(q-1)}  e^{i q\Phi})]^* V_{ref}^2 +P_{cor} \nonumber \\
                &  = &  g^{2}G_{1}G_{2}^* [1- | \Psi | ^{2}(\displaystyle\sum\limits_{p=1}^{\infty}  \Upsilon^{(p-1)}  e^{i p\Phi})(\displaystyle\sum\limits_{q=1}^{\infty}  \Upsilon^{(q-1)}  e^{i q\Phi})^{*} \nonumber \\
  & & + 2 i {\rm Im} \{ \Psi (\displaystyle\sum\limits_{m=1}^{\infty}  \Upsilon^{(m-1)}  e^{i m\Phi}) \} ] V_{ref}^2 + P_{cor}.
  \end{eqnarray}
 As before, the responses in switch positions `0' and `1' have the relationship: $(P_{1ref} - P_{cor}) = -(P_{0ref} - P_{cor})$. Subtracting the cross-power in the two switch positions to cancel $P_{cor}$, we get
\begin{eqnarray} 
P_{ref} & = &  P_{1ref} - P_{0ref} \nonumber \\
  &=& -2 g^{2}G_{1}G_{2}^* V_{ref}^2  \nonumber \\
  && +2 g^2 G_1 G_2^* | \Psi | ^{2} \displaystyle\sum\limits_{m=1}^{\infty}  |\Upsilon^{m-1}|^{2} [1+ {\rm Re} (\Upsilon e^{i\Phi}) + {\rm Re} (\Upsilon^2 e^{i2\Phi})  + \nonumber \\ 
  && ........ +{\rm Re} (\Upsilon^{n-1} e^{i (n-1) \Phi}) + ........ ] V_{ref}^2 \nonumber \\
&&  + i 4 g^2 G_1 G_2^* {\rm Im} \{ \Psi (\displaystyle\sum\limits_{m=1}^{\infty}  \Upsilon^{m-1}  e^{i m\Phi}) \}  V_{ref}^2. 
  \end{eqnarray}

Compared to the response to antenna temperature, the response to reference noise is complex.  The response to multi-path propagation of reference noise is less than that due to multi-path propagation of antenna noise owing to the appearance of an additional factor $\Psi$. This factor $\Psi = (\Gamma_{1}-\Gamma_{2})\Gamma_{a}g^{2}$ has a magnitude of about 2\% in the SARAS system.  

The first term in Equation~49 represents the nominal system response neglecting multi-path propagation, which is the response to reference noise traveling directly to the pair of LNAs in the two arms of the correlation spectrometer.  When calibrated, this term corresponds to the ambient temperature of the matched termination provided by the reference, which is about 300~K.

The second term is a `real' response that is similar in mathematical description to the series of terms in Equation~42.  However, because of the term $| \Psi |^2$ that appears as a common multiplier, all terms corresponding to differential delays exceeding $2l$ may be neglected since the response to these are expected to be less than 1~mK.  Additionally, only the $m=1$ term in the expansion need be considered. The `real' response will, therefore, have a spectral ripple with amplitude less than 6~mK and the ripple will have the fundamental period corresponding to a differential delay of $2 l$; harmonics will have amplitudes well below 1~mK.

The third term in Equation~49 represents the `imaginary' response.  This is terms corresponding to correlation products 
between the direct propagation of reference noise along one arm with the $n$'th reflected component along the other path.   These products  represent terms of type $d_{1}r_{2}^*$ and $r_{1}d_{2}^*$ and the net correlation is imaginary.  The imaginary response is a Fourier series of ripples with zero expectation for the mean response.   In the SARAS system, the fundamental is expected to have amplitude about 6~K; harmonics have their amplitudes reduced progressively by factors about 5\% and, therefore, we may limit consideration to just the first and second harmonics while modeling the spectra to 1~mK accuracy.  

Retaining the significant terms reduces Equation~49 to
\begin{eqnarray} 
P_{ref} & \approx & -2 g^{2}G_{1}G_{2}^* [ 1 -  | \Psi | ^{2} \{ 1+ {\rm Re} (\Upsilon e^{i\Phi}) \} -2i {\rm Im} ( \Psi e^{i \Phi} + \Psi \Upsilon e^{i 2\Phi} \nonumber \\
 & & + \Psi \Upsilon^2 e^{i 3\Phi}) ] V_{ref}^2. 
\end{eqnarray}

\subsubsection{SARAS response to receiver noise}

Components of the receiver noise traveling from the LNA inputs opposite to the signal flow path encounter the antenna and reference terminations.  As discussed earlier we neglect the reflections from the reference because of the excellent impedance match provided at that port and consider only the dominant reflections from the antenna terminal.  In this section, we develop the expressions for the net  complex response of the SARAS correlation spectrometer to receiver noise; this analysis considers multiple reflections of the LNA noise between the amplifier inputs and antenna terminals.   When the cross-over switch connects the antenna to the $\Sigma$ port of the power splitter the sum of  the upward traveling noise voltages from the two LNAs reaches the antenna and when the antenna is connected to the $\Delta$ port of the power splitter the difference between the voltages propagates to the antenna. Following similar notations as in previous sections and considering the state of the system with cross-over switch in position `0', the net signal voltages at the pair of amplifier inputs are:
 \begin{eqnarray} 
   V_{1} & =  &  V_{n1}+ (f_{1}V_{n1} - f_{2}V_{n2}) \Gamma_a g^2 e^{i\Phi}+
   (f_{1}V_{n1} - f_{2}V_{n2})(\Gamma_1+ \Gamma_2) (\Gamma_a g^2)^{2}e^{i2\Phi} \nonumber \\
   && +  ........ +  (f_{1}V_{n1} - f_{2}V_{n2})(\Gamma_1+ \Gamma_2)^{(n-1)} (\Gamma_a g^2)^n e^{in\Phi} + ......\nonumber\\
     & = & V_{n1} + (f_{1} V_{n1} - f_{2} V_{n2}) \displaystyle\sum\limits_{m=1}^{\infty} (\Gamma_a g^2)^{m}(\Gamma_1+ \Gamma_2)^{(m-1)}  e^{i m\Phi} 
 \end{eqnarray}
   and, similarly,
  \begin{eqnarray}
   V_{2} & =  &  V_{n2} - (f_{1} V_{n1} - f_{2} V_{n2}) \displaystyle\sum\limits_{m=1}^{\infty} (\Gamma_a g^2)^{m}(\Gamma_1+ \Gamma_2)^{(m-1)}  e^{i m\Phi}.
 \end{eqnarray}
The complex response of the correlation spectrometer is:
\begin{eqnarray} 
P_{0n} & = & (G_{1}V_{1}) \times (G_{2}V_{2})^* + P_{cor} \nonumber  \\
&=& G_1 G_2^* [ V_{n1} V_{n2}^* + (f_{1} V_{n1} - f_{2} V_{n2}) V_{n2}^* \chi - (f_{1} V_{n1} - f_{2} V_{n2})^* V_{n1} \chi^* \nonumber \\
&& - (f_{1} V_{n1} - f_{2} V_{n2})(f_{1} V_{n1} - f_{2} V_{n2})^* \chi \chi^* ] + P_{cor},
\end{eqnarray}
where we have used the symbol $\chi$ to denote the series expansion:
\begin{eqnarray}
\chi & = & \displaystyle\sum\limits_{m=1}^{\infty} (\Gamma_a g^2)^{m}(\Gamma_1+ \Gamma_2)^{(m-1)}  e^{i m\Phi} \nonumber \\
& = & g^2 \Gamma_a e^{i \Phi}  \displaystyle\sum\limits_{m=1}^{\infty} \Upsilon^{m-1} e^{i (m-1) \Phi}.
\end{eqnarray}

The time integral of the product of noise voltages from the two LNAs is expected to vanish; therefore, the SARAS response only contains the terms
\begin{eqnarray} 
P_{0n} & = & - G_{1}G_{2}^* [V_{n1}^2 (f_1 \chi^* + f_1^2 \chi \chi^*) + V_{n2}^2 (f_2 \chi + f_2^2 \chi \chi^*)] + P_{cor}.
\end{eqnarray}

With cross-over switch in position `1', a similar analysis leads to the response
\begin{eqnarray} 
P_{1n} & = &  G_{1}G_{2}^* [V_{n1}^2 (f_1 \chi^* + f_1^2 \chi \chi^*) + V_{n2}^2 (f_2 \chi + f_2^2 \chi \chi^*)] + P_{cor}.
\end{eqnarray}

Differencing the measured complex responses in the two switch positions,
\begin{eqnarray}
P_{n} & = & P_{1n}-P_{0n} \nonumber \\
& = & 2G_{1}G_{2}^* [V_{n1}^2 (f_1 \chi^* + f_1^2 \chi \chi^*) + V_{n2}^2 (f_2 \chi + f_2^2 \chi \chi^*)] \nonumber \\
& = & 2G_{1}G_{2}^* [V_{n1}^2 f_1 \chi^* + V_{n2}^2 f_2 \chi + (V_{n1}^2 f_1^2 + V_{n2}^2 f_2^2) \chi \chi^* ] \nonumber \\
& = & 2 g^2 G_{1}G_{2}^* f_1 V_{n1}^2 \Gamma_a^* e^{-i\Phi} (\displaystyle\sum\limits_{m=1}^{\infty} \Upsilon^{m-1} e^{i (m-1) \Phi})^* \nonumber \\
 & &   + 2 g^2 G_{1}G_{2}^* f_2 V_{n2}^2 \Gamma_a e^{i\Phi} \displaystyle\sum\limits_{m=1}^{\infty} \Upsilon^{m-1} e^{i (m-1) \Phi} \nonumber \\
 & & + 2 g^2 G_{1}G_{2}^* f_1^2 V_{n1}^2 |\Gamma_a|^2 (\displaystyle\sum\limits_{p=1}^{\infty} \Upsilon^{p-1} e^{i (p-1) \Phi}) (\displaystyle\sum\limits_{q=1}^{\infty} \Upsilon^{q-1} e^{i (q-1) \Phi})^* \nonumber \\
 & & + 2 g^2 G_{1}G_{2}^*  f_2^2 V_{n2}^2 |\Gamma_a|^2 (\displaystyle\sum\limits_{p=1}^{\infty} \Upsilon^{p-1} e^{i (p-1) \Phi}) (\displaystyle\sum\limits_{q=1}^{\infty} \Upsilon^{q-1} e^{i (q-1) \Phi})^*.
\end{eqnarray}

The first line in the above expansion represents correlation products between the receiver noise ($V_{n1}$) of the LNA in Arm~1 with the component of this receiver noise ($f_1 V_{n1}$) that propagates in the reverse direction from the input port of that LNA and arrives at the correlator along Arm~2.  This component may incur multiple reflections between the LNA inputs and antenna and hence the response would be a harmonic series of ripples in the spectral response.  The second line in the above expansion represents a similar response for the receiver noise of the LNA in Arm~2.  The third line represents correlation products between the reverse propagating receiver noise from the LNA in Arm~1, which arrives at the correlator along both arms after multiple reflections between the antenna and LNA terminals; these signals have relative delays of zero and multiples of $2 l$.  The fourth line is a similar response to reverse propagating receiver noise from the LNA in Arm~2.

We adopt parameters relevant to the SARAS system, where $f_1$ and $f_2$ are in the range 5--10\%, and the magnitudes of $\Upsilon$ and $\Gamma_a$ are, respectively below 5 and 10\%.  The receiver noise temperatures are about 115~K.  

The responses written as the third and fourth lines in the above Equation~57 contain terms that arise from receiver noise that propagates along the two arms after multiple reflections at the antenna terminals.  This response is `real' and with magnitude less than 0.01\% of the receiver temperature, or about 10~mK.  Only terms corresponding to singly-reflected receiver noise propagating along the two arms are significant.  These are the terms corresponding to $p=1$ and $q=1$ and for these terms the response is relatively a smooth function of frequency.
 
All ripples in the response that are above mK in amplitude arise from the first two lines of the expansion.  The ripple with fundamental period corresponding to path delay of $2 l$ is expected to have an amplitude of less than one per cent of the receiver temperature, or less than 1~K.  Successive harmonics have amplitudes reduced by $|\Upsilon|$, which is less than 5\%.  Therefore, we may consider the first harmonic alone while modeling observations for EoR signatures.

Retaining these significant terms, the SARAS response to receiver noise may be approximated to:
\begin{eqnarray}
P_{n} & \approx & 2 g^2 G_{1}G_{2}^* [ (f_1^2 V_{n1}^2 +  f_2^2 V_{n2}^2)  |\Gamma_a|^2 + \nonumber \\
 & & (f_1 V_{n1}^2 +  f_2 V_{n2}^2) \times  {\rm Re} \{ \Gamma_a ( e^{i \Phi} + \Upsilon e^{i 2 \Phi}) \} + \nonumber \\
 & & i(f_1 V_{n1}^2 -  f_2 V_{n2}^2) \times  {\rm Im} \{ \Gamma_a ( e^{i \Phi} + \Upsilon e^{i 2 \Phi}) \} ].
 \end{eqnarray}
If the two LNAs form a matched pair then the imaginary component of the response would vanish.  In SARAS, we expect a 3:1 ratio between the amplitudes of the sinusoidal response in the real and imaginary components.

\subsection{Bandpass Calibration revisited}

The primary calibration of the complex gain of the correlation spectrometer is based on the calibration noise source, which is switched on and off in each of the switch positions.  This noise source is coupled to the signal path via a directional coupler, and the switching of the noise power is achieved by turning the noise source on and off rather than switching the coupling of the power to the signal path.  Both these approaches are aimed at maintaining the impedance of the system interconnects unchanged during the calibration cycle, so that the calibration signal suffers the same reflections within the system as the antenna signal, and the calibration may be used to calibrate the spectral response to the antenna signal including effects of multi-path propagation.

Following the same notation as in Section~3.2, we denote the responses of the system when the calibration noise source is in `on' state as $P_{1cal}$ and $P_{0cal}$ in the two states of the cross-over switch.   The responses when the calibration noise source is `off' are $P_{1off}$ and $P_{0off}$.  

The net response of the system is:
\begin{eqnarray}
P_{off} & = & P_{1off} - P_{0off} \nonumber \\
& = & (P_{1a} - P_{0a}) + (P_{1ref} - P_{0ref}) + (P_{1n} - P_{0n}) \nonumber \\
& = & (P_{a} + P_{ref} + P_{n})
\end{eqnarray}

The double difference:
\begin{eqnarray}
P_{bpc} & = & P_{cal} - P_{off} \nonumber\\
               & = & (P_{1cal} - P_{0cal}) - (P_{1off} - P_{0off})
\end{eqnarray}
represents the response of the system to calibration noise alone and has the same form as the response to antenna noise power: the response is given by Equation~42 except that the antenna signal $V_{a}^2$ is replaced by $V_{cal}^2$.  The response to calibration noise is wholly `real'.    The bandpass calibration is
\begin{eqnarray}
P_{bpc} & = & 2g^{2}G_{1}G_{2}^* [\displaystyle\sum\limits_{m=0}^{\infty}(|\Upsilon^{m}|^{2}) 
\displaystyle\sum\limits_{k=0}^{\infty} {\rm Re} (\Upsilon^k e^{ik\Phi})] V_{cal}^2. \nonumber \\
&=&  2g^{2}G_{1}G_{2}^* [\displaystyle\sum\limits_{m=0}^{\infty}(|\Upsilon^{m}|^{2}) ] V_{cal}^2 + \nonumber \\
&& 2g^{2}G_{1}G_{2}^* [ {\rm Re} (\Upsilon e^{i\Phi}) \displaystyle\sum\limits_{m=0}^{\infty}(|\Upsilon^{m}|^{2}) ] V_{cal}^2 + \nonumber \\
&& 2g^{2}G_{1}G_{2}^* [ {\rm Re} (\Upsilon^2 e^{i2\Phi}) \displaystyle\sum\limits_{m=0}^{\infty}(|\Upsilon^{m}|^{2}) ] V_{cal}^2 + \nonumber \\
&& ........... + 2g^{2}G_{1}G_{2}^* [ {\rm Re} (\Upsilon^n e^{in\Phi}) \displaystyle\sum\limits_{m=0}^{\infty}(|\Upsilon^{m}|^{2}) ] V_{cal}^2 + ...........
\end{eqnarray}

Bandpass calibration of the net response is complex arithmetic and involves dividing the net response of the system (Equation~59) with the above $P_{bpc}$ so that the complex frequency dependent gain $2g^2 G_1 G_2^*$ is calibrated for.  Additionally, the measured spectrum is also given a temperature scale by multiplying the quotient with the strength of the calibration noise $T_{cal}$ in temperature units:
\begin{eqnarray}
T_{a}^{\prime} & = & {(P_{a} + P_{ref} + P_{n})\over P_{bpc} }\times T_{cal} 
 = T_a + [ {{P_{ref} + P_n}\over{P_{bpc}}} ] T_{cal}.
\end{eqnarray}

The bandpass calibration yields a complex result.  The real part is the sum of the antenna temperature spectrum and bandpass calibrated $(P_{ref} + P_n)$.  The SARAS system requires a joint modeling of the real and imaginary parts of $T_a^\prime$ to separate the contributions from reference noise and receiver noise and isolate $T_a$.  Useful in this modeling is the expectation that $T_a$ has no contribution to the imaginary part of $T_a^\prime$.  

In Table~\ref{table:param} we provide a summary of the quantities that must be solved for.  

\section{Modeling of SARAS data} 

\begin{figure}
\centering
\includegraphics[width=11.5cm]{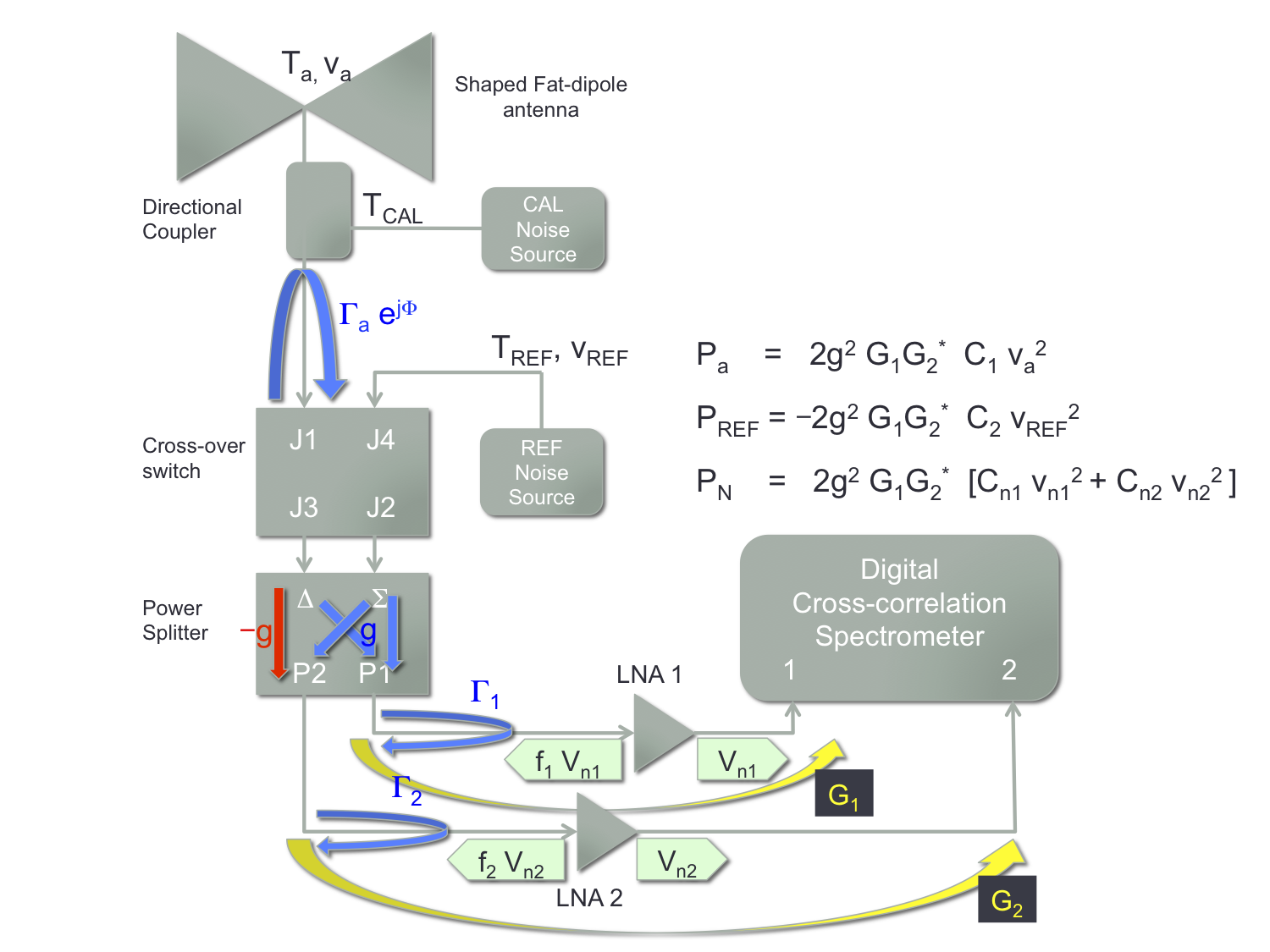}
\caption{Figure depicts the various gains, reflections coefficients and noise powers/voltages that constitute the model for the SARAS spectrometer.} 
\label{fig:design2}       
\end{figure}

In this final section, we discuss the data products from the spectrometer and an approach to modeling the system towards deriving useful measurements of the cosmic radio background and of re-ionization signatures embedded therein.  Fig.~\ref{fig:design2} shows a diagram of the SARAS spectrometer depicting the noise powers and corresponding signals that are of significance: noise powers from the antenna ($P_{a}$), reference source ($P_{ref}$) and the receiver ($P_{n}$) and corresponding voltages $V_a$, $V_{ref}$ and $V_{n1}$ and $V_{n2}$.  Also depicted are the gains $g$, $G_1$ and $G_2$ within the power splitter and in the two arms containing the receiver chains, the voltage reflection coefficients $\Gamma_{a}$, $\Gamma_{1}$ and $\Gamma_{2}$ at the antenna terminals and at the two LNA inputs, and the noise voltages originating in the LNAs in the forward ($V_{n1}$ and $V_{n2}$) and reverse ($f_1 V_{n1}$ and $f_2 V_{n2}$) directions.  The figure also shows the equations for the responses of the system in terms of $C_{1}$, $C_{2}$, $C_{n1}$ and $C_{n2}$,  which are functions of  many of the above system parameters depicted here.  

\subsection{The SARAS measurement set}

During an observation, SARAS measures two complex bandpass calibrated cross-power spectra. One of them is a differential measurement of the sky temperature with reference to a `cold' reference load and the other is the difference between the sky temperature and a `hot'  reference.  We have used the terms `hot' and `cold' reference loads to refer to the on and off states of the reference noise source.  The receiver noise also contributes to the differential output measurement  due to cross coupling between the receiver channels, which is primarily a consequence of the mismatch at the antenna terminals.  As discussed previously, in these differential measurements the sky power (antenna temperature) appears only in the real part.   Following on from Equation~62, the above differential measurements ($T_{a0}^{\prime}$ and $T_{a1}^{\prime}$ respectively) may be written in temperature units as:
\begin{equation}
T_{a0}^{\prime}  = G_a T_{sky} -   {C_{2} \over C_{1}} T_{ref0}
 +{C_{n1} \over C_{1}}  T_{n1} 
 +{C_{n2} \over C_{1}} T_{n2}
\label{ref_cold}
\end{equation}
and
\begin{equation}
T_{a1}^{\prime}  = G_{a} T_{sky} -   {C_{2} \over C_{1}} T_{ref1}
 +{C_{n1} \over C_{1}}  T_{n1} 
 +{C_{n2} \over C_{1}} T_{n2},
\label{ref_hot}
\end{equation}
where $T_{a}$, $T_{ref0}$, $T_{ref1}$, $T_{n1}$ and $T_{n2}$ are the antenna temperature, hot and cold reference load temperatures 
and the two receiver noise temperatures respectively.  $G_a$ is the antenna gain, which is a function of frequency and defines the antenna bandpass.  The frequency response of the antenna includes the change in gain of the fat dipole versus frequency as well as the frequency dependence in the response to sky temperature that arises from the ground reflection; both these are multiplicative and modeled as a single antenna gain term $G_a$. The primary data product or measurement set from SARAS is a series of  $T_{a0}^{'}$ and $T_{a1}^{'}$ complex spectra in the frequency range 87.5 to 175~MHz with a spectral resolution of 85.4~kHz; a pair is computed every 4.2~s.  The SARAS antenna has a frequency independent pattern that is a dipole; the axis of the dipole is parallel to the ground and hence half the pattern is on the sky and the other half sees an absorber-covered ground.

From the measurement set additional calibration products are derived: one dependent only on the reference temperature step and the other on differential sky temperature. The first is obtained by differencing the two measurements made with the reference termination in the `hot' and `cold' states, {\it i.e.}, by differencing Equations~63 \& 64 at each integration time to get
\begin{equation}
T_{a0}^{'} - T_{a1}^{' } =   {C_{2} \over C_{1}} (T_{ref1}- T_{ref0}).
\label{dif_ref}
\end{equation}
To derive the second, we difference sky spectra measured at different local sidereal times.  For example, a particular realization of this differential sky spectrum may be obtained by dividing the measurement set into two equal halves based on the median antenna temperature during the entire observation,  the measurements with above-median antenna temperatures form the `hot' sky spectra and the other half forms the `cold' sky spectra. We may write the mean antenna temperatures over `hot' and `cold' skies as $T_{ah} = G_{a} T_{hot~sky}$ and  $T_{ac} = G_{a} T_{cold~sky}$ respectively in terms of the mean sky brightness temperatures over the `hot' and `cold' skies.  We compute the difference between the hot and cold sky spectra to get
\begin{equation}
T_{ah}^{'} - T_{ac}^{' } =    T_{ah} - T_{ac} = G_a (T_{hot~sky} - T_{cold~sky}).
\label{dif_sky}
\end{equation}
The first calibration product is a measure of the response to reference load, including calibration.  The second yields calibration products that serve for joint modeling of the parameters for the foreground sky model and those of $G_a$, the frequency response of the antenna or its bandpass shape.

In Table~\ref{table:param}, which contains a summary of the quantities that must be solved for, we point out the dependencies of the real and imaginary components of the measurement set (Equation~63) and also the dependencies of the calibration products given by Equation~65 and 66.

\begin{table}
\caption{A summary of the parameters that describe the system and must be determined.  The table also shows the dependencies of the real and imaginary components of the measurement set and derived calibration products on these parameters: a 
$\surd$ symbol indicates dependence. The data products also have different dependencies on the spectra of the Galactic anisotropic component and the extragalactic (cosmological) sky brightness; these are also shown in the Table.}
\label{table:param}
\centering
\begin{tabular}{|c|c|c|c|c|c|c|}
\hline
 \multicolumn{2}{|c|}{ Parameters}   & \multicolumn{2}{|c|}{$T_{a0}^{'}$} & \multicolumn{2}{|c|}{$T_{a0}^{'}-T_{a1}^{'}$} 
 & {$T_{ah}-T_{ac}$}    \\
 \multicolumn{2}{|c|}{ } &  \multicolumn{2}{|c|}{ Re~~Im } &  \multicolumn{2}{|c|}{Re~~~~Im} &   \\ 
\hline
 & $G_{a}$    & $\surd $ & - &  -  &  - & $\surd$    \\
\cline{2-7}
& $\Gamma a$ & $\surd $ & $\surd $ & -   & -   & -    \\
\cline{2-7}
System & $\tau $   & $\surd $ & $\surd $ & $\surd $ & $\surd $ & -   \\
\cline{2-7}
& $\Upsilon=(\Gamma_{1}+\Gamma_{2})\Gamma_{a} g^{2}$  & $\surd $ & $\surd $  & $\surd $ & $\surd $ & -   \\
\cline{2-7}
& $\Psi=(\Gamma_{1}-\Gamma_{2})\Gamma_{a} g^{2}$ & $\surd $ & $\surd $ &  $\surd $ & $\surd $ & -    \\
\cline{2-7}
 & $T_{ref0}$    & $\surd $ & $\surd$ & - & - & -   \\
\cline{2-7}
 & $T_{ref1} - T_{ref0}$   & - & - & $\surd$  & $\surd$ & -    \\
\cline{2-7}
 & $f_{1}^{2} T_{n1} + f_{2}^{2} T_{n2}$   &$\surd $ & - & - & -   & -   \\
\cline{2-7}
  & $f_{1} T_{n1} + f_{2} T_{n2}$    & $\surd $ & - & - & -  & -   \\
\cline{2-7}
 & $f_{1} T_{n1} - f_{2} T_{n2}$    & - & $\surd$ & - & -   & -    \\
\cline{2-7}
\hline
Sky & Anisotropic sky  & $\surd $ & - & - & - & $\surd $   \\
\cline{2-7}
Spectrum & Isotropic sky & $\surd $ & - & - & - & -  \\
\hline
\end{tabular}
\end{table}

Critical to bandpass calibration of the measurement is the calibration noise source, which is injected into the signal path via the directional coupler at the antenna.  The calibration of the equivalent noise temperature and flatness of this noise source is part of the observing strategy; this involves replacing the antenna with a matched load with temperature sensor and thermally cycling the termination and analyzing the measurement set to solve for the characteristics of the calibration noise source.  Unfortunately at the precision of the SARAS system, the complex impedance of terminations appear to change with temperature, and this manifests as a scaling and shift in the measured ripples arising from reflections of reference noise and receiver noise at the termination.   The process of primary calibration includes solving for this shift and scale in the imaginary part of the spectrum and correcting for this effect.

\subsection{Instrument model}

To the degree of complexity that we have considered in our analysis presented here, the SARAS system is described in terms of eight system characteristics. The measurement set may be related to the sky spectrum via the measurement equations, which depends on the gain of the antenna ($G_{a}$) and that of the power splitter ($g$), complex reflection coefficients at the antenna terminals ($\Gamma_{a}$) and at the inputs of the low-noise amplifiers
($\Gamma_{1}$ \& $\Gamma_{2}$), the time delay in the propagation path between the antenna and receiver ($\tau = {\Phi}/(2\pi \nu)$), and the fractional LNA noise voltage that propagates out the LNA input terminals ($f_{1}$ \& $f_{2}$).  Of these, $\tau$, $g$, $f_{1}$ and $f_{2}$ are modeled as  independent of frequency.   In expanded form the defining model for the instrument is in Equations~42,  49 \& 57. 

As discussed in previous sections, the SARAS system is configured in experimental setups to measure $\Gamma_{a}$, $\Gamma_{1}$, $\Gamma_{2}$, $f_{1}$ and $f_{2}$.  Antenna measurements yield the absolute scaling factor $G_{a}$ at the band center; an estimate for $\tau$ is obtained from the length and characteristics of the coaxial cable interconnect between the antenna and receiver.    All of these yield good initial guesses for the system parameters, and fits to these measurements provide fitting forms whose parameters may be refined in the modeling of the data.

\subsection{Sky model}

The antenna temperature $T_{a}$ is the product of sky temperature $T_{sky}$ and the antenna gain-loss function $G_{a}$.   Therefore modeling the antenna temperature involves modeling the sky spectrum and $G_{a}$. 

At any instant $T_{sky}$ is the average of the sky brightness temperature weighted by the antenna beam; the sky brightness along any line of sight is the cosmic radio foreground ($T_{fg}$) due to extragalactic radio sources and the Galactic synchrotron emission in addition to redshifted 21~cm background ($T_{bg}$) from cosmological distribution of neutral hydrogen gas.  The foregrounds are modeled using a functional form similar to the one used by Pritchard and Loeb (2010). As we have shown in Section~3.4, to reduce the modeling residuals to mK levels the beam-averaged sky spectrum requires modeling with four parameters and using a third order polynomial in log-temperature versus log-frequency space.   The foreground sky spectrum is described using the form
\begin{equation} 
{\rm log}_{10} T_{fg} = a_{0} + a_{1} ({\rm log}_{10}( \frac{\nu}{\nu_{o}}) )+ a_{2} ({\rm log}_{10}(\frac{\nu}{\nu_{o}}))^2 + a_{3} ({\rm log}_{10}(\frac{\nu}{\nu_{o}}))^3,
\end{equation}
where $\nu_{o}$=131.25~MHz is the mid frequency of our band. 

The simplest model for the background reionization feature that may be expected imprinted on the cosmic radio background is a toy model `step'.  This has a tanh-form discontinuity that is parameterized by the redshift of reionization $z_{r}$ and the duration of reionization $\Delta_{z}$ (this model was adopted in the WMAP7 analysis; Larson et al. (2010)):
 \begin{equation}
 T_{bg}(z) = {T_{21}\over 2} (\sqrt{(1+z) \over 10} [ {\rm tanh} ({(z-z_{r})\over \Delta_{z}})+1].
 \end{equation}
 $T_{21}$ is the amplitude of the reionization signal.  More realistic models for reionization lead to expectations of the forms derived in Pritchard and Loeb (2008) and in Pritchard and Loeb (2010), whose detection requires bandwidths in excess of an octave and ability to discern broad spectral signatures in the presence of orders of magnitude more intense continuum foregrounds.

\subsection{Modeling the primary and the derived spectra: a hierarchical approach} 

The primary measurement is Equation~\ref{ref_cold}, in which the antenna temperature appears in the real component. The imaginary part of Equation~\ref{ref_cold} and the derived calibration products described in Equations~\ref{dif_ref} \& \ref{dif_sky} do not contain features of the uniform cosmological radio background and reionization signatures: the modeling of these spectra calibrate the instrument parameters.

The first step in the modeling of the measurement set is the modeling of the difference spectrum computed between `hot' and `cold' reference, which is described in Equation~\ref{dif_ref}.  This depends on parameters $\Upsilon = (\Gamma_1+ \Gamma_2) \Gamma_a g^2$ and $\Psi = (\Gamma_1- \Gamma_2) \Gamma_a g^2$, whose complex reflection coefficients are measured in the laboratory except for the gain term $g$.  This also depends on the propagation delay $\tau$ and fitting for $\tau$ is expedited by having a long cable length for the interconnect between the antenna and receiver, which would cause a number of cycles of ripple in the spectrum.  Such an approach is preferred when low order background spectral features are sought to be detected as, for example, in measurements of the absolute spectrum of the cosmic radio background.  However, in observations aiming to model the data for plausible `steps' arising from cosmological reionization, a short interconnect is preferred.

This step of the modeling, which has a focus on $\Upsilon$ and $\Psi$, underscores a positive aspect of the SARAS design.  The modeling at this stage is of the ratio of parameters $C_2$ and $C_1$.  These parameters include ripples across frequency space of form $\Upsilon e^{i\Phi}$ and $\Psi e^{i\Phi}$, which are modulated by the very terms $\Upsilon$ and $\Psi$ that are also parts of additive functions in $C_2$ and $C_1$ that manifest as low order spectral features.  The design has the advantage of allowing solutions for the low order terms via solving for the modulations in the high order ripples in the spectrum, which may be introduced by choice of cable length to be of a high order and disjoint in Fourier space from astrophysical features.

The next step in the hierarchical modeling is to expand the modeling to include the imaginary component of the primary measurement Equation~\ref{ref_cold}.  This step introduces instrument parameters related to receiver noise, this second step adds the term $(f_{1} V_{n1}^2 - f_{2} V_{n2}^2)$ to the instrument model. 

The last step in this hierarchical modeling is to include the real component of the primary measurement together with the calibration product that is the difference between `hot' and `cold' sky: Equation~\ref{dif_sky}.  Additional instrument parameters related to receiver noise are now included in the model: $(f_{1} V_{n1}^2 + f_{2} V_{n2}^2)$ and $(f_{1}^2 V_{n1}^2 + f_{2}^2 V_{n2}^2)$.  The relative gain of the antenna versus frequency---$G_a$---is another parameter that is introduced into the solver at this stage and is constrained by the difference Equation~\ref{dif_sky}.

\begin{figure}
\centering
\includegraphics[width=11.5cm]{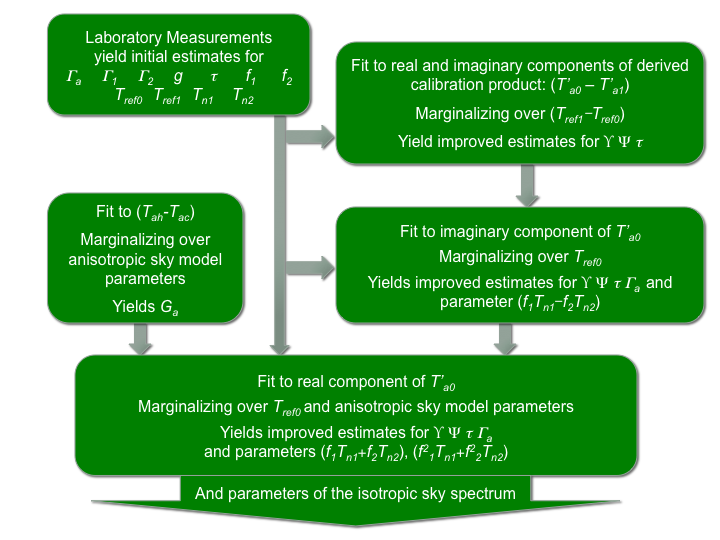}
\caption{Figure depicts the various gains, reflections coefficients and noise powers/voltages that constitute the model for the SARAS spectrometer.} 
\label{fig:flowchart}       
\end{figure}

A flowchart depicting the steps in the modeling of the parameters that describe the system as well as the sky spectrum is shown in Fig.~\ref{fig:flowchart}.

The SARAS system design thus allows a hierarchical approach where the terms defining the instrument model may be solved for first in the imaginary component of the measurement set and in calibration products, which do not have any response to spectral features in the cosmic radio background, before moving on to the final stages of modeling wherein terms defining the sky model are brought into the solver.  Modeling at all stages may adopt optimization algorithms in which all the unknown parameters are constrained simultaneously within a Bayesian framework of analysis.  The approach also allows for the parameters describing the reflection coefficients $\Gamma_a$, $\Gamma_1$ and $\Gamma_2$ to be refined in the solver.  A detailed discussion on the Monte Carlo Markov chain (MCMC) as a methodology for estimation of cosmological parameters and simultaneous solution for instrument coefficients can be found in Harker et. al. (2012). 

\section{Postlude}
 
 Analysis of SARAS measurements for deriving constraints on parameters defining the sky spectrum and its features requires an analysis pipeline that derives likelihood estimates for sky model parameters given measurement sets.  The next step in SARAS is development and implementation of such a code that operates on simulated data sets that include the non-ideal instrument behavior described in this work.
 
 We have designed SARAS with a $180^{\circ}$ hybrid as a power splitter for the correlation receiver.  The current implementation uses non-identical low-noise amplifiers.  The analysis presented herein shows that using a matched pair of amplifiers cancels certain contaminants. In the response to reference noise and receiver noise the imaginary components are canceled and, therefore, the net response of the system is now purely real.  The real component is also altered in that this component of the response to reference noise is now ideal.  The advantage of using matched amplifiers in the SARAS configuration is that the contamination in the real component of the calibrated spectrum, which contains the antenna temperature, is marginally reduced.  However, the downside is that the calibration product in the form of the imaginary component of the response is lost. On balance, we have preferred to implement the system without using matched amplifiers because the additional contamination that this results in is expected to be only of a few mK amplitude.
 
The system configuration could have used a quadrature hybrid that is a $90^{\circ}$ power splitter in place of the $180^{\circ}$ hybrid.  Differencing the measurements in the two states of the cross-over switch would continue to cancel the correlator-based additive noise.  In the case of a quadrature hybrid, the response of the system would manifest the antenna and reference temperatures in the imaginary component instead of the real component.  However, after calibration using the response to the calibration noise, the difference between antenna and reference temperatures would, as in the case of the $180^{\circ}$ hybrid, appear in the real component.  
 
The primary advantage of using a quadrature hybrid is in the substantial reduction of reflection back to the input port when matched amplifiers are used.  This reduces the ripple in the response to antenna temperature because of the reduced reflection of this power back from the amplifiers to the antenna.  The reference noise power, which dominates the ripple in the calibrated spectrum in the case of a $180^{\circ}$ hybrid, will continue to appear as the dominant ripple in the calibrated measurement set.  Additionally, ripples owing to multi-path propagation of receiver noise will also continue to appear. The advantage of using a quadrature hybrid is that higher order ripples arising from multiple reflections between the antenna and amplifiers are reduced.   The analysis presented above will be simplified, and higher order terms derived here may be neglected, if a quadrature hybrid along with a matched pair of low-noise amplifiers replaces the power splitter in future.  The alternate method of suppressing reflections from the low-noise amplifier inputs, while retaining a $180^{\circ}$ hybrid, is to implement each of the amplifier modules as a combination of a a quadrature hybrid followed by a pair of matched amplifiers followed by a second quadrature hybrid.  

The SARAS design presented herein has adopted a $180^{\circ}$ hybrid followed by wide-band amplifiers without any quadrature hybrid. This choice was motivated by practicalities in that the performance of available wideband $180^{\circ}$ hybrids is superior to that of quadrature hybrids.  In so far as we know, $180^{\circ}$ hybrids can be made to be wideband and with smooth characteristics because they contain simple transformers; however, quadrature hybrids require a high order transfer function that will have correspondingly high order characteristics.  If this is placed in the top of the receiver chain the modeling of the internal system will have more parameters to solve for.  The real answer to whether a quadrature hybrid or $180^{\circ}$ hybrid ought to be used, and whether matched amplifiers are preferred in the SARAS configuration, will come from modeling using an analysis pipeline that derives likelihood estimates for sky model parameters given measurement sets.  This may separately analyze the alternate configurations considering real-world models for the components to ascertain the structure for which the EoR parameters are better constrained in their likelihood estimates; as mentioned above this is part of future work in the SARAS development.

 \section{Summary}
 
 SARAS is a ground-based experiment for precision measurements of the cosmic radio background at low frequencies over an octave band from 87.5 to 175~MHz.  Measurement of a wideband sky spectrum at these long wavelengths with the exacting requirements demanded for detecting any faint signatures corresponding to redshifted 21-cm from the epoch of reionization motivated the development of a new design concept including a calibration strategy, and understanding the non-ideal behavior of the system to the extent that the science goals demand.  
 
 The antenna in the SARAS configuration is a frequency-independent Fat-dipole over an absorber ground plane.  We argue for a correlation spectrometer with the power splitter preceded by a cross-over switch, which acts as a differential radiometer providing measurements of the difference between antenna temperature and noise temperature of an internal reference termination.  The calibration noise is injected at the antenna via a directional coupler so that the interconnects and system behavior are unaltered during calibration.    Analog signal transmission from the compact antenna electronics to a base station at a remote location is via an optic fiber link; this was another essential design feature that provided exceptional reverse isolation in the signal flow path and hence avoids unwanted coupling between the signals in the two arms of the correlation receiver.
 
 We provide a detailed analysis of the signal processing including inevitable effects related to multi-path propagation within the receiver electronics.  We discuss coupling paths for the signal flow and the propagation of noise from the antenna, from the reference termination and from the low-noise amplifiers.  We provide a formalism that describes the effects of multiple internal reflections between the impedance mismatches encountered by the signals, expanding the reflections as infinite series.  We use measurements of reflection coefficients of sample devices, as well as of a frequency-independent Fat-dipole antenna, to derive example responses to antenna noise, reference termination and amplifier noise.  These serve as a guide to the complexity necessitated of the modeling that derives constraints on parameters that describe sky spectral features.  
 
A key advantage of the SARAS configuration is that the measurement set is in the form of complex spectra with sky power appearing wholly in one component, the real component.  Spectral structure arising from multi-path propagation due to internal reflections appears in both components and, most importantly, shares parameters between the real and imaginary components.  The system design also provides derived calibration products that are once again devoid of sky temperature and provide data that constrain the parameters describing the spectral contaminants.   

This paper has described the system design and all of the rationale in the system design.  The analysis of the signal flow and processing lays the foundation for the use of the system for measurements of the cosmic radio background spectrum.  Varying the length of the interconnecting cable between the antenna and receiver varies the delay $\tau$ in the above analysis and moves the responses arising from multi-path propagation into different Fourier components of the measured spectrum; thus SARAS measurements of the sky radio spectrum with differing lengths for $l$ would be preferred for different science.  SARAS is being deployed for absolute and precision measurements of the spectrum of the cosmic radio background in the 87.5 to 175~MHz band and for spectral features that may arise from hydrogen reionization in the redshift window 7.1-15.2.

\begin{acknowledgements}

The development of the SARAS concept has benefited greatly from the contemporaneous experimentation with implementations and trials of a variety of system configurations, which received outstanding support from staff of the Gauribidanur Radio Observatory and the Electronics Laboratory and Mechanical Engineering services at the Raman Research Institute.  We thank Ron Ekers for his encouragement and participation in the effort.  

\end{acknowledgements}

\end{document}